\documentclass[structabstract]{aa} 
\usepackage[fleqn]{amsmath}
\usepackage[varg]{txfonts}
\usepackage{graphicx,txfonts, ulem}
\usepackage{natbib}
\usepackage{hyperref}
\bibpunct{(}{)}{;}{a}{}{,}
\begin{document}
 \title{r-Java 2.0: the nuclear physics}

      \author{M. Kostka\thanks{email:mkostka@ucalgary.ca}
          \inst{1}
		\and
          N. Koning
          \inst{1}
          \and
          Z. Shand 
          \inst{1}          
          \and
          R. Ouyed 
          \inst{1}
          \and
          P. Jaikumar
          \inst{2}
          }

   \institute{Department of Physics \& Astronomy, University of Calgary, 
2500 University Drive NW, Calgary, Alberta, T2N 1N4 Canada             
		\and Department of Physics \& Astronomy, California State University Long Beach, 1250 Bellflower Blvd., Long Beach, California 90840 USA         
        }

   \date{Received ; accepted }

  \abstract
   {}
   {We present r-Java 2.0, a nucleosynthesis code for open use that performs r-process calculations as well as a suite of other analysis tools. }
   {Equipped with a straightforward graphical user interface, r-Java 2.0 is capable of; simulating nuclear statistical equilibrium (NSE), calculating r-process abundances for a wide range of input parameters and astrophysical environments, computing the mass fragmentation from neutron-induced fission as well as the study of individual nucleosynthesis processes.}
   {In this paper we discuss enhancements made to this version of r-Java, paramount of which is the ability to solve the full reaction network.  The sophisticated fission methodology incorporated into r-Java 2.0 which includes three fission channels (beta-delayed, neutron-induced and spontaneous fission) as well as computation of the mass fragmentation is compared to the upper limit on mass fission approximation.  The effects of including beta-delayed neutron emission on r-process yield is studied.  The role of coulomb interactions in NSE abundances is shown to be significant, supporting previous findings.  A comparative analysis was undertaken during the development of r-Java 2.0 whereby we reproduced the results found in literature from three other r-process codes.  This code is capable of simulating the physical environment of; the high-entropy wind around a proto-neutron star, the ejecta from a neutron star merger or the relativistic ejecta from a quark nova.  As well the users of r-Java 2.0 are given the freedom to define a custom environment.  This software provides an even platform for comparison of different proposed r-process sites and is available for download from the website of the Quark-Nova Project: \url{http://quarknova.ucalgary.ca/}   }
   {}


   \maketitle
   
   \authorrunning{Kostka et al.}
	\titlerunning{R-process Nucleosynthesis with r-Java 2.0}
\section{Introduction}
\label{intro} 
A key nucleosynthesis mechanism for the production of heavy elements beyond the iron peak is rapid neutron capture, or the r-process \citep{burb57,cam57}.  In spite of a large volume of observational data \citep{sneden08}, the astrophysical site(s) of the r-process remains an open question.  Explosive, neutron-rich environments provide the ideal conditions for r-process to occur.  The predominant astrophysical sites which are being studied as possible locations for r-process are: the high-entropy winds (HEW) from proto-neutron stars \citep[see][]{qianWoos96,farouqi10} and ejected matter from neutron star mergers \citep[see][]{fried99,gori11}.  However both of these scenarios face significant hurdles that must be overcome.  The long time-scale for neutron star merger events \citep{faber12} limits this scenario's ability to explain the r-process element enrichment of metal-poor stars \citep{sned03}.  HEW models have been shown to be very sensitive to the chosen physical conditions of the winds and elaborate hydrodynamic models have yet to prove that the HEW scenario can provide the necessary environment for significant r-process to occur \citep[e.g.,][]{hoff08,janka08,rob10,fisc10,wanajo11}.  The theoretical quark nova has as well been proposed as a potential r-process site \citep{jaik07}.  The explosive, neutron-rich environment of a quark nova provides an intriguing avenue of study for astrophysical r-process \citep{jaik07}.  It is important to note that the observed nuclear abundance of r-process elements in metal-poor stars \citep{sned03} and abundance data of certain radionuclides found in meteorites \citep{qian08} point to the likelihood of multiple r-process sites \citep{tru02}.  Certainly there remains much to be learned about astrophysical r-process and its necessary conditions.  To help drive this study we have developed r-Java \citep{rjava1}, which is a cross-platform, flexible r-process code that is transparent and freely available for download by any interested party\footnote{r-Java 1.0 \& 2.0 can be downloaded from quarknova.ucalgary.ca}.

The purpose of this article is to introduce the second version of r-Java (r-Java 2.0), discuss the new features which it contains and display a selection of simulation results.  Prior to delving into the details of r-Java 2.0 it would be enlightening to briefly examine the capabilities and limitations of the first version of the code.  As discussed in \cite{rjava1} our aim for r-Java was to create an easy-to-use, cross-platform r-process code that avoids the $^{\backprime}$black-box$^{\prime}$ pitfalls that plague many scientific codes.  In order to achieve this goal, r-Java was developed with an intuitive graphical user interface (GUI) and we provide extensive documentation as well as user tutorials on our website. The first version of r-Java was predicated upon the waiting point approximation (WPA) which assumes an equilibrium between neutron captures and photo-dissociations.  The practical effect of using the WPA is that the relative abundance along isotopic chains depends only on neutron density ($n_n$), temperature ($T$) and neutron separation energy ($S_n$), \citep[For more details see][eqn. 12]{rjava1}.  Through the WPA the number of coupled differential equations that must be solved is reduced from thousands to on the order of a hundred (the number of isotopic chains in the reaction network), greatly reducing computational costs.  The utilization of the WPA comes at the expense of generality, in that the assumption is only valid in high temperature and neutron density environments (typically only considered for $T_9 > 2$ ,where $T_9$ is in units of $10^9$ K and $n_n > 10^{20}$ cm$^{-3}$).  Within the context of the WPA r-Java 1.0 provided users the ability to run r-process simulations for a wide variety of scenarios, including neutron irradiation of static targets as well as the dynamic expansion of r-process sites.  A key feature of r-Java that has been maintained through the release of the second version is the ability for users to easily make changes to nuclear inputs between simulation runs.  Other flexibilities of r-Java include the ability to specify the amount of heating from neutrinos, turn on/off various processes and choose the density profile of the expanding material.

This paper is organized as follows.  Section \ref{code} provides an overview of the advancements made in r-Java 2.0 and discusses the default rates included with the code.  The nuclear statistical equilibrium (NSE) module is discussed in section \ref{nse}.  The new fission methodology is detailed in section \ref{fiss}.  The effect of $\beta$-delayed neutron emission is discusssed in section \ref{bdn}.  Section \ref{full} compares the full reaction network calculation with the WPA approach.  Comparative analysis between r-Java 2.0 and other full reaction network codes is carried out in section \ref{bench}.  Finally in section \ref{summary} a summary is provided along with a look to future work on r-Java.

\section{Overview of r-Java 2.0}
\label{code}

\begin{figure*}
\includegraphics[width = \linewidth]{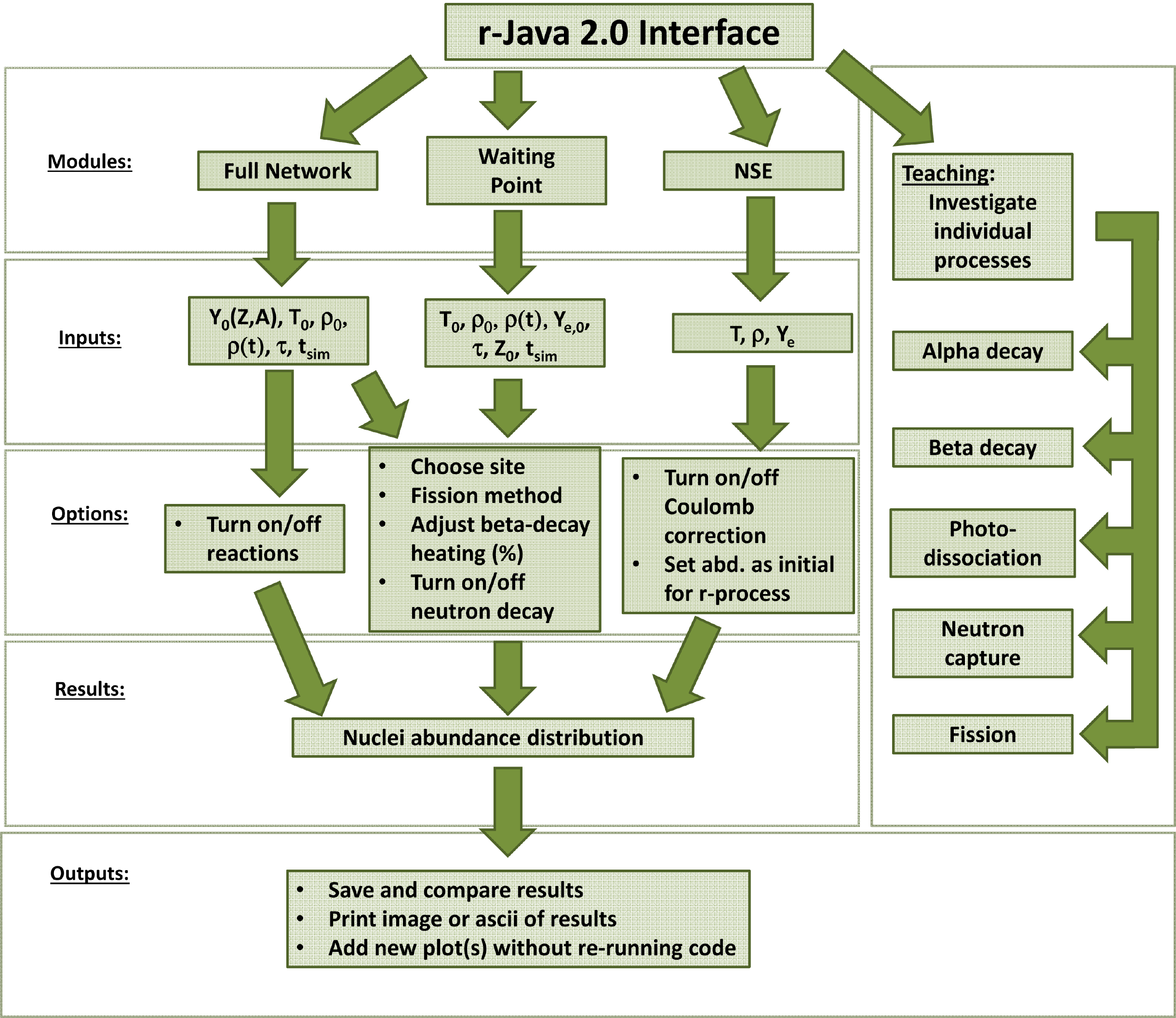}
\caption{A schematic representation of the functionality of r-Java 2.0.  See Table \ref{symbol} for description of symbols.  }
\label{flow}
\end{figure*}

There are major developments that have been made to r-Java since the original release in 2011.  Most note-worthy, r-Java 2.0 is now capable of solving a full reaction network and is no longer solely reliant on the WPA.  The latest release also contains a more accurate handling of fission as well as the implementation of $\beta$-delayed neutron emission of up to three neutrons.  Another expansion to r-Java is the ability for the user to specify the astrophysical environment of the r-process, which determines the methodology used to evolve the density and temperature.  Further, the nuclear statistical equilibrium (NSE) module has been expanded to include the effect of coulomb screening.  Finally any nuclear reaction can be turned on or off which allows the user to investigate individual processes.  An organizational chart displaying the functionality of r-Java 2.0 can be seen in Fig. \ref{flow}.

As seen in Fig. \ref{flow}, r-Java 2.0 contains several distinct modules.  NSE, WPA network and full network constitute the research modules which are aimed to be used by scientists in order to study r-process.  Complementary components to r-Java 2.0 are the teaching modules, aimed for use in the classroom at the graduate and undergraduate levels, these modules allow for investigation of individual nucleosynthesis processes.  The fission module calculates the mass fragmentation of neutron-induced fission.  The user is able to choose the target nucleus (or nuclei), vary the incident neutron energy as well as adjust four parameters related to the potential energy of the fragmentation channels (to be discussed in section \ref{fiss}).  The remaining teaching modules; $\beta$-decay, $\alpha$-decay, photo-disocciation and neutron capture all act in similar manners.  The user can specify an initial abundance of nuclei and then investigate how varying the rates affects the final abundances for different physical conditions.    

\subsection{Rates}
\label{ratesSect}
The default rates and cross-sections are based on the Hartree-Fock-Bogolubov 21 (HFB21) mass model\citep{HFBmass} as calculated by the reaction code {\sc TALYS} \citep{TALYS}.  The publicly available Maxwellian-averaged neutron capture cross-sections and corresponding photo-dissociation rates are provided on a temperature grid that extends from $10^6$ K to $10^{10}$ K \citep{TALYS}.  Because neutron capture cross-sections and photo-dissociation rates can change by many orders of magnitude between temperature grid points a simple cubic spline interpolation was insufficient.  A unique interpolation method was developed that does not fall victim to the over-shooting and correction of a normal cubic spline.  To avoid adding uncertainty by extrapolating the photo-dissociation rates and neutron capture cross-sections, the extremal values of the temperature grid provide the temperature bounds for the r-process calculations in r-Java 2.0.

\begin{figure*}
\includegraphics[width = \linewidth]{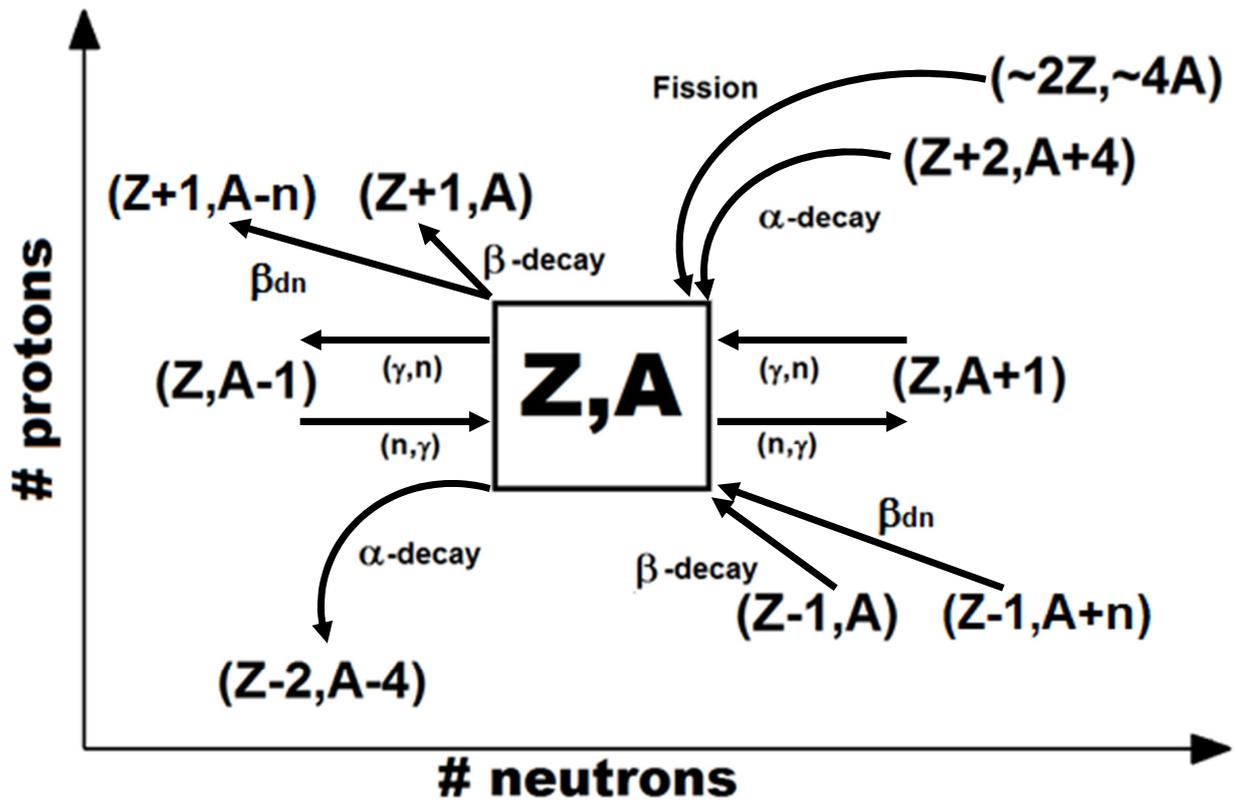}
\caption{The reactions incorporated in the full reaction network calculation in r-Java 2.0 are represented schematically for a given isotope (Z,A).  These reactions are: neutron-capture (n,$\gamma$), photo-dissociation ($\gamma$, n), $\beta$-decay, beta-delayed neutron emission ($\beta$dn), $\alpha$-decay and fission.  }
\label{schemeRates}
\end{figure*}

The $\beta^-$ decay half-lives and probability of $\beta$-delayed neutron emission of up to three neutrons are considered in r-Java 2.0 by making use of the calculations done by \cite{moller03}.  For complete consistency these rates should be calculated using the HFB21 mass model, however to our knowledge such a calculation has yet to be carried out.  Alpha decay half-lives are calculated based on an empirical formula dependant on the ejected alpha particle kinetic energy \citep{lang80}.  Fig. \ref{schemeRates} shows a schematic representation of all the processes that are incorporated in the full reaction network calculation. 

\begin{figure*}
\includegraphics[width = \linewidth]{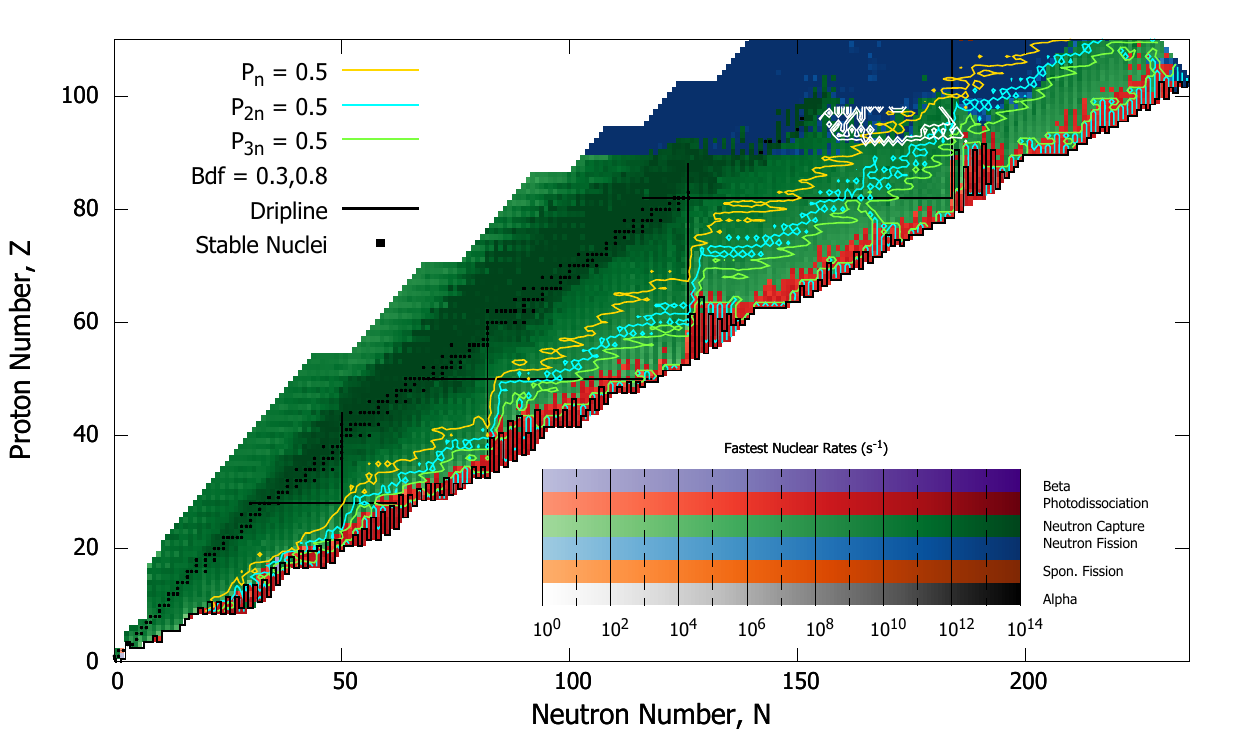}
\caption{For each nuclei in our network the fastest rate is plotted given a temperature of $1 \times 10^9$K and a neutron density of $1\times 10^{30}$cm$^{-3}$.  The contour lines indicate when the probability of $\beta$-delayed emission of \textit{n} neutrons reaches 50\%.  The neutron drip line and the location of the proton and neutron magic numbers are denoted with black solid lines. The location of the stable nuclei are denoted by the black squares.  A colour version of this figure is available in the online article. }
\label{hinnlot}
\end{figure*}

\begin{figure*}
\includegraphics[width = \linewidth]{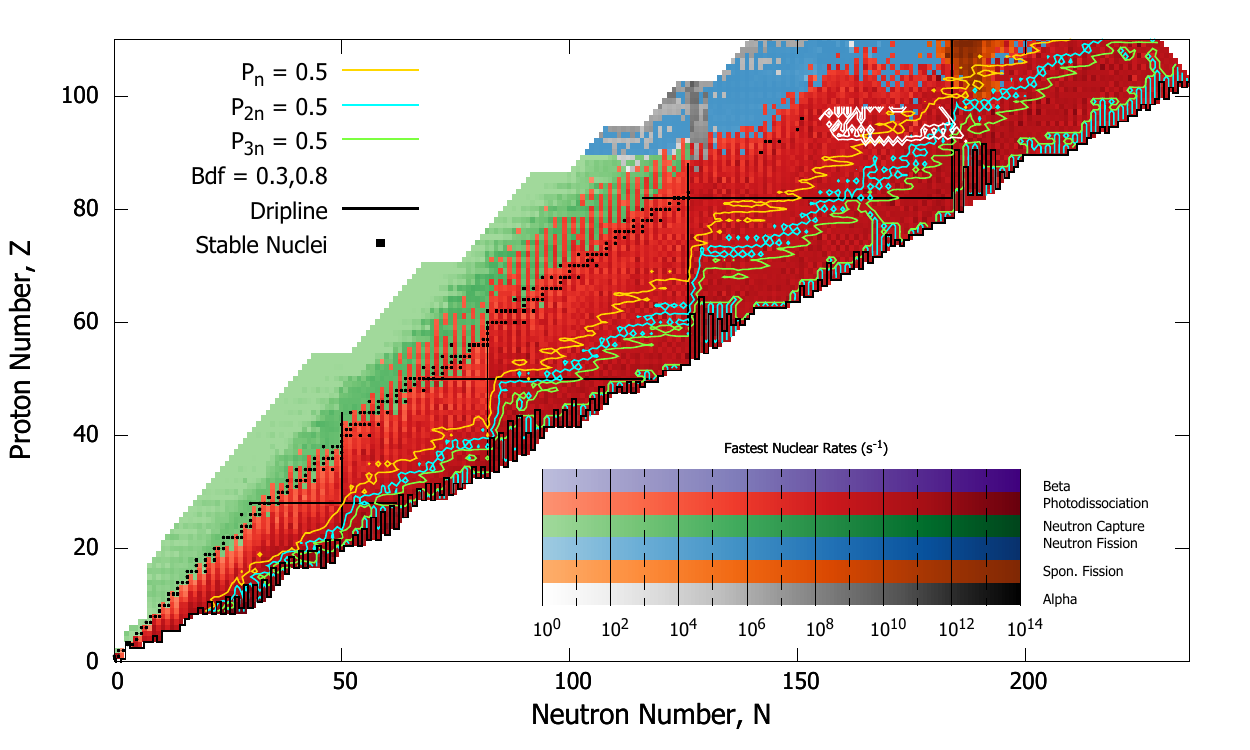}
\caption{Same as Fig. \ref{hinnlot} but withn a temperature of $3 \times 10^9$K and a neutron density of $1\times 10^{20}$cm$^{-3}$.  A colour version of this figure is available in the online article. }
\label{lonnhit}
\end{figure*}

\begin{figure*}
\includegraphics[width = \linewidth]{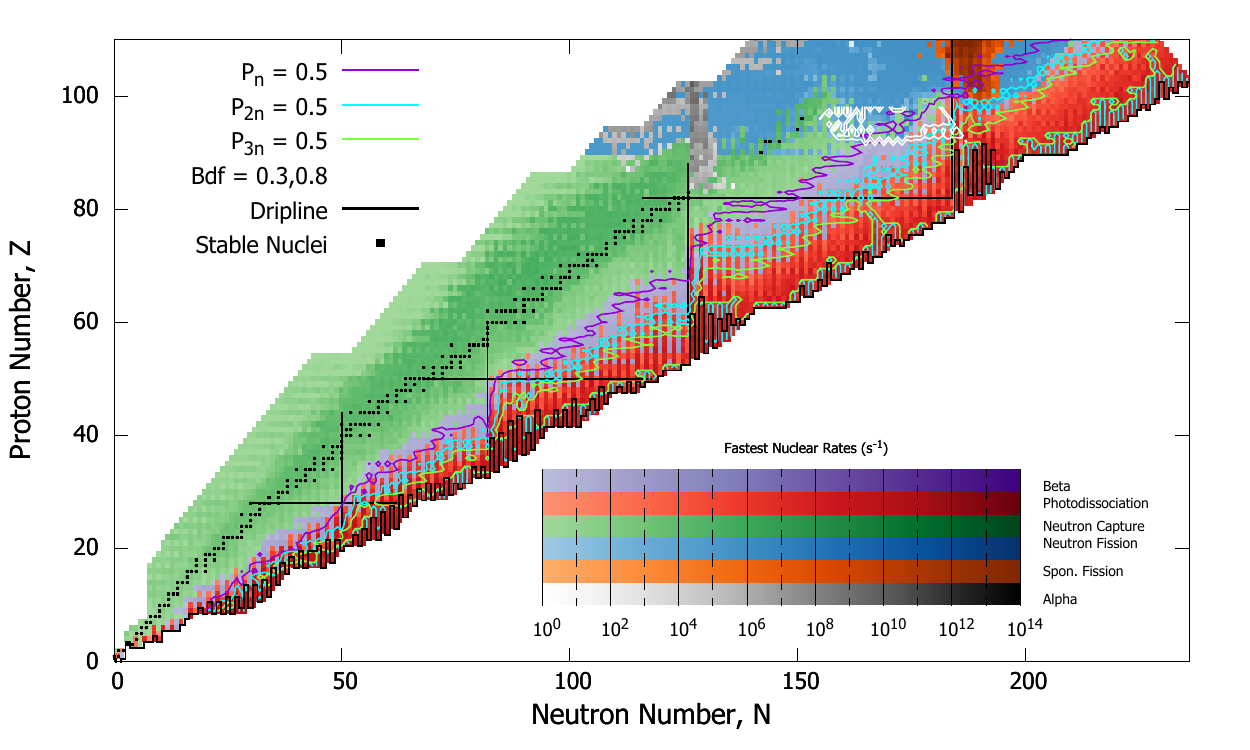}
\caption{Same as Fig. \ref{hinnlot} but withn a temperature of $1 \times 10^9$K and a neutron density of $1\times 10^{20}$cm$^{-3}$.  A colour version of this figure is available in the online article.}
\label{lonnlot}
\end{figure*}

Since r-Java 2.0 makes use of temperature dependant neutron capture cross-sections, photo-dissociation rates and neutron-induced fission cross-sections at any given temperature and neutron density the dominant transmutational process for each nuclei could be different.  Figs. \ref{hinnlot}, \ref{lonnhit} and \ref{lonnlot} consider all available processes in r-Java 2.0 and display the dominant process for each nuclei in our network at three different neutron density and temperature combinations; fig \ref{hinnlot} - ($\log \left( n_{\rm n} \right) = 30, T_9 = 1$), fig \ref{lonnlot} -  ($\log \left( n_{\rm n} \right) = 20, T_9 = 1$) and fig \ref{lonnhit} -  ($\log \left( n_{\rm n} \right) = 20, T_9 = 3$).  For the high neutron density, low temperature scenario shown in Fig. \ref{hinnlot} neutron capture or neutron-induced fission are the dominant channels for most nuclei in the network.  Photo-dissociation is the strongest rate only for the most neutron-rich isotopes of each element. The waiting points at the N = 82, 126 and 184 closed shells can be seen as steps along the interface between neutron capture and photo-dissociation the latter being the dominant process.  In the high temperature, low neutron density case displayed in Fig. \ref{lonnhit} photo-dissociation becomes the most probable channel for the majority of nuclei in the network.  $\alpha$-decay dominates neutron-induced fission for some neutron-poor isotopes of heavy elements, in particular in a region around the N = 126 closed shell.  For the low neutron density, low temperature example shown in Fig. \ref{lonnlot} $\beta$-decay is the dominant channel for a band of nuclei that stretches nearly the entire length of the network.  Odd-even effects can be seen as along many isotopic chains $\beta$-decay and photo-dissociation alternate as the dominant process.  The region in which $\alpha$-decay   dominates is similar to but more robust than the high temperature low neutron density case seen in Fig. \ref{lonnhit}.  The fertile and fissile regime is dominated by neutron-induced fission save a region in which spontaneous fission is the dominant decay channel, this region of spontaneous fission instability can also be seen in Fig. \ref{lonnhit}.  

A fundamental tenet followed during the development of r-Java 2.0 was maximizing the flexibility afforded to the user.  To this end, built into the r-Java 2.0 interface is a module dedicated to displaying and editing the nuclear parameters.  The user of r-Java 2.0 can modify any parameter (mass or $\beta$-decay rate for instance) in between simulation runs, without having to restart the program.  This allows users of r-Java 2.0 to quickly and easily test the effect of changing nuclear properties on r-process abundances.  The choice of Java as the language to develop our nucleosynthesis code was made to ensure that r-Java 2.0 could be used across all platforms.  Special attention was paid to designing a graphical user interface that is intuitive and easy-to-use.

\subsection{Getting Started with r-Java 2.0}
\label{started}

As discussed above, maximizing flexibility was paramount when developing r-Java 2.0.  This extends beyond the nuclear inputs and to the astrophysical parameters which govern how the temperature and density of the system evolve.  With r-Java 2.0 we endeavoured to create r-process software that could be applied to any potential astrophysical r-process site.  For this purpose the user of r-Java 2.0 can choose from a set of astrophysical sites which provide unique density evolutions and related input parameters.  The choices for astrophysical sites are; high-entropy winds around a proto-neutron star, ejecta from a neutron star merger or the ejecta from a quark nova.  The details of the specific physics implemented for each of the astrophysical sites  will be discussed in a forthcoming paper.  If one chooses to go beyond the aforementioned astrophysical sites a custom density evolution can be selected.  The user of r-Java 2.0 is free to define any dynamical evolution for the density or choose a static r-process site.  For the remainder of this work we will be considering custom density evolution equations. 

With the nuclear and physical parameters chosen, before an r-process simulation can be run the user must determine the initial abundances of the r-process site.  This is handled differently for the WPA and full network modules.  When entering a WPA simulation the user may specify the initial electron fraction (Y$_{\rm e}$) and element, then based on this information and the initial temperature r-Java 2.0 computes the starting neutron density ($n_{\rm n,0}$) and isotopic abundances using Maxwell-Boltzmann statistics.  If the user chooses a full network simulation the initial mass fractions must be specified after which r-Java 2.0 calculates the starting Y$_{\rm e,0}$ and $n_{\rm n,0}$ ensuring that baryon number and charge are conserved.  

\begin{figure*}
\includegraphics[scale = 0.8]{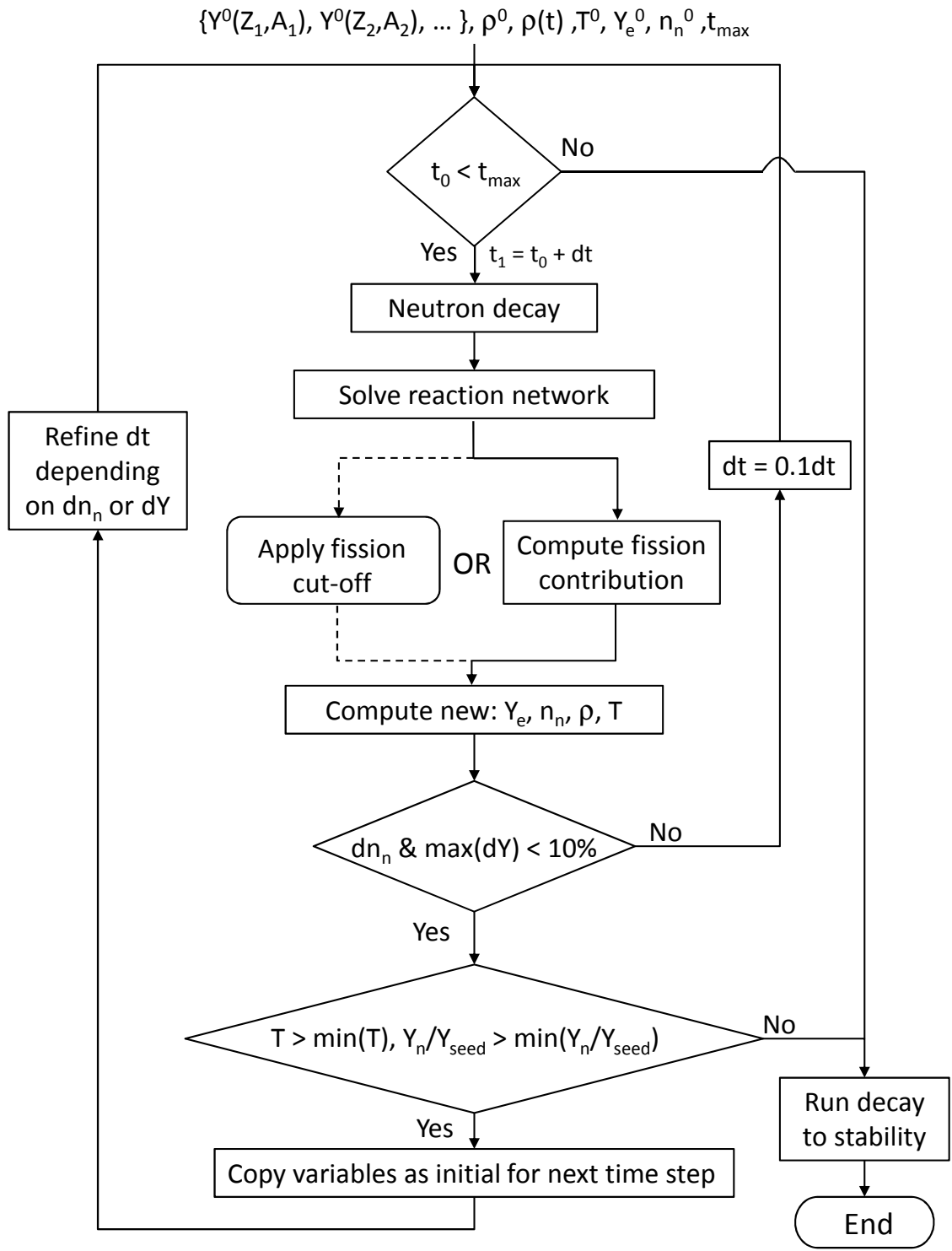}
\caption{A schematic representation of a single time-step in our full network code.  $\lbrace Y_0(Z_1,A_1), Y_0(Z_2,A_2), ...\rbrace$ denotes the set of initial nuclei abundances, $\rho_0$ is the initial mass density, $\rho(t)$ defines the density evolution, $T_0$ is the initial temperature, $Y_{\rm e,0}$ denotes the initial electron fraction and $n_{\rm n,0}$ is the initial neutron density.  First the neutron decay is computed before the reaction network is solved using the Crank-Nicholson algorithm.  Next the fission contribution is calculated along with the new physical parameters.  If the changes in abundance or $n_{\rm n}$ are too large the time-step is reattempted with $dt$ = 0.1$dt$. Adaptive time-steps are used in order to maximize $dt$.  }
\label{chart}
\end{figure*}

Once the initial condition are determined the r-process code follows the algorithm detailed in Fig. \ref{chart} and runs until the user-specified duration is met or one of the stopping criterion is satisfied. The minimum temperature stopping criterion is determined by the neutron capture and photo-dissociation temperature grid.

\section{Nuclear Statistical Equilibrium}
\label{nse}

This release of r-Java includes a refinement to the NSE module whereby the user can now choose to include the effect of Coulomb screening.  Under NSE the nuclei abundances are uniquely determined by three parameters; Y$_{\rm e}$, mass density ($\rho$) and T.  In the most conventional sense, when a system that follows Maxwell-Boltzmann statistics is said to be in NSE the particle number density of nuclei \textit{i} which contains Z protons and N neutrons (where mass number A = Z + N) is given by \citep[e.g.][]{pathria77}

\begin{equation}
n_i = g_i \left(\frac{2\,\pi\,k\,T}{h^2}\right)^{3/2}\rm{exp}\left(\frac{\mu_i +B_i}{k\,T}\right)
\end{equation}

\noindent where $T$ represents the temperature of the system, $k$ is Boltzmann's constant, $h$ is Planck's constant and $B_i$, $\mu_i$, $g_i$ denote the binding energy, chemical potential and statistical weight respectively.  When the Coulomb correction is applied, $\mu_{\rm C,tot}$ is added in the exponential where

\begin{equation}
\mu_{\rm C,tot} = Z\,\mu_{\rm C, p} -\mu_{\rm C}(Z,A).
\end{equation}

\noindent This correction to the chemical potential arises from the Coulomb contribution to the free energy which becomes significant for heavier nuclei ($\mu_{\rm C, p}$ is the Coulomb potential of a bare proton).  Our methodology for calculating $\mu_{\rm C}(Z,A)$ is similar to that of \cite{gori11} and is given by 
\begin{equation}
\mu_{\rm C}(Z,A) = k\,T\,f_{\rm C}(\Gamma_i)
\end{equation}

\noindent where $f_{\rm C}(\Gamma_i)$ is the Coulomb free energy per ion in units of $k\,T$.  For a Coulomb liquid $f_{\rm C}(\Gamma_i)$ can be expressed as \citep{haensel07}

\begin{equation}
\begin{split}
f_{\rm C}(\Gamma_i) = \\ A_1\,\sqrt{\Gamma_i \left(A_2+\Gamma_i \right) } \\ - A_1 \times A_2 {\rm ln}\left(  \sqrt{\Gamma_i / A_2}+\sqrt{1+\Gamma_i / A_2}  \right) \\ + 2A_3 \left(  \sqrt{\Gamma_i} -{\rm arctan}\left(  \sqrt{\Gamma_i}  \right) \right) \\ + B_1 \left(  \Gamma_i - B_2 {\rm ln} \left(  1 + \frac{\Gamma_i}{B_2}  \right)  \right)  +  \frac{B_3}{2}{\rm ln}\left( 1 + \frac{\Gamma_i^2}{B_4} \right)
\end{split}
\end{equation}

\noindent with $A_1 = -0.9070, A_2 = 0.62954, A_3 = 0.27710, B_1 = 0.00456, B_2 = 211.6, B_3 = -0.0001, B_4 = 0.00462$.  When a user chooses to include the Coulomb correction r-Java 2.0 will only do so if the Coulomb liquid approximation is valid, which is to say that the Coulomb coupling parameter, 

\begin{equation}
\Gamma_i = \frac{Z^2\,e^2}{a_i\, k\, T}
\end{equation}

\noindent where $a_i$ is the ion-sphere radius, is smaller than the melting value $\Gamma_{\rm m} = 175.0 \pm 0.4$ \citep{potekhin00}.  

\begin{figure*}

\includegraphics[width = \linewidth]{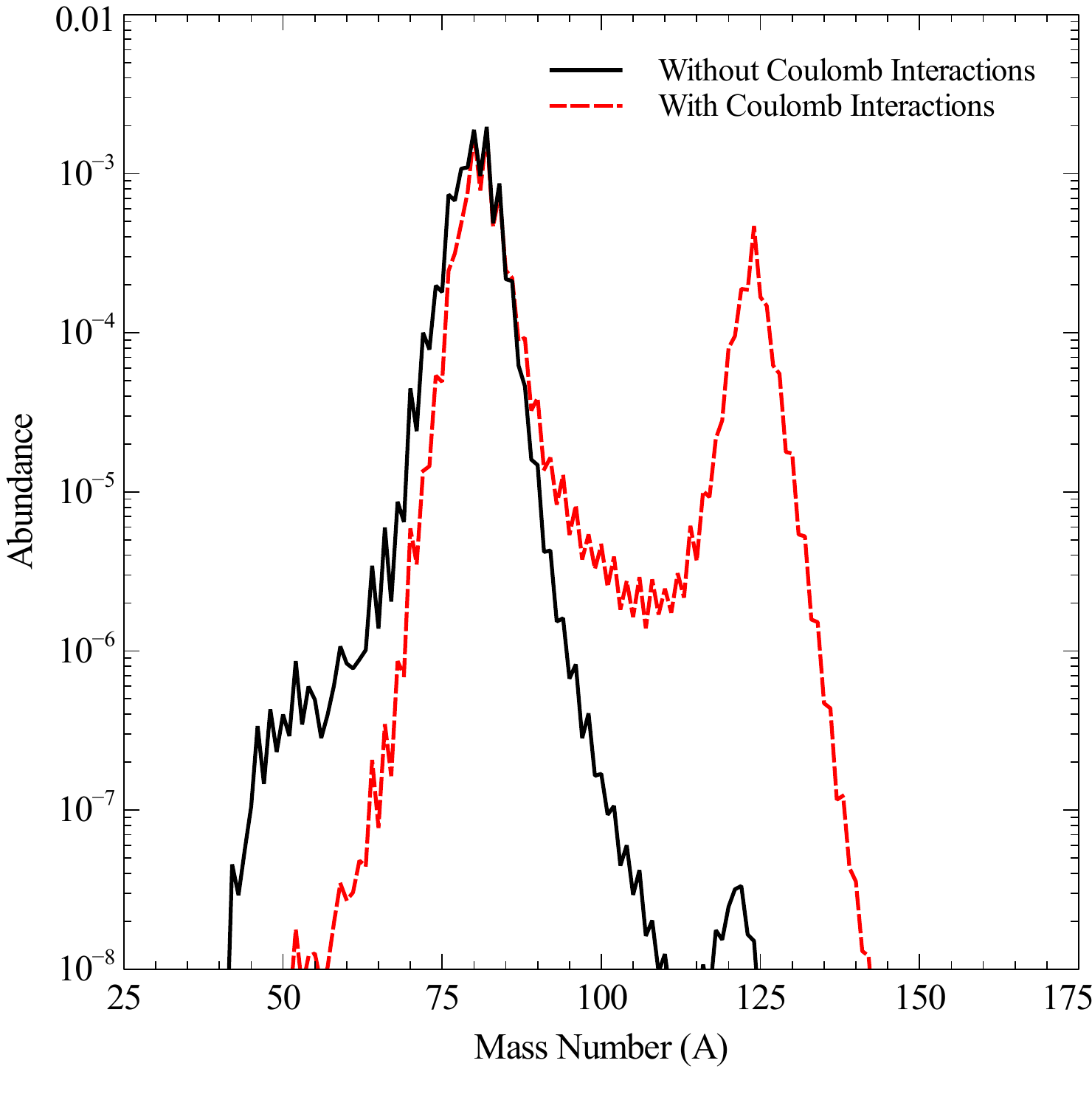}
\caption{NSE abundance distribution subject to the following physical conditions; temperature of 1$\times$10$^{10}$K, mass density of 2$\times$10$^{11}$g$\,$cm$^{-3}$ and electron fraction of 0.3.  The red dashed line denotes a calculation that includes the effects of Coulomb interactions while for the black solid line the Coulomb interactions were ignored. }
\label{nseFig}
\end{figure*}

The effect of including Coulomb screening can be seen in Fig. \ref{nseFig} which displays an overlay of two NSE abundances both considering the same temperature ($T=1\times10^{10}$K), mass density ($\rho = 2 \times 10 ^{11}$g cm$^{-3}$) and electron fraction ($Y_{\rm e} =0.3$), the only difference being whether Coulomb screening is included.  Coulomb screening can allow for the formation of a significant amount of heavier elements.  As the example in Fig. \ref{nseFig} shows, with the Coulomb correction to the chemical potential included a peak appears at approximately A = 124 that is absent in the case where Coulomb screening is ignored.

Since r-process requires an explosive astrophysical site there is a likelihood that the material that will undergo r-process will have begun in NSE \citep{gori11}.  To accommodate such r-process scenarios r-Java 2.0 gives the user the option to run the NSE module and set the resulting nuclei abundance as the initial abundance for an r-process simulation.  Currently under development is a charged particle reaction network module that will be incorporated in a future release of r-Java.

\section{Fission}
\label{fiss}
The previous release of r-Java instituted a simple maximum Z and A approach to fission.  After the reaction network was solved species with larger values of Z or A than the imposed limit were split into two smaller species \citep[see][for more details]{rjava1}.  For r-Java 2.0 the users are given the option to turn off fission, choose the same cut-off approach as in the previous r-Java release or choose a more realistic treatment that includes spontaneous, neutron-induced and $\beta$-delayed fission.  Spontaneous fission rates are computed using the logic presented by \cite{kodoma75} and the $\beta$-delayed fission probabilities were taken from \cite{panov05}.  Fission barrier heights and neutron-induced fission rates provided as defaults in r-Java 2.0 are calculated by \cite{gori09} based on the HFB14 mass model.   

\begin{figure*}
\includegraphics[width = \linewidth]{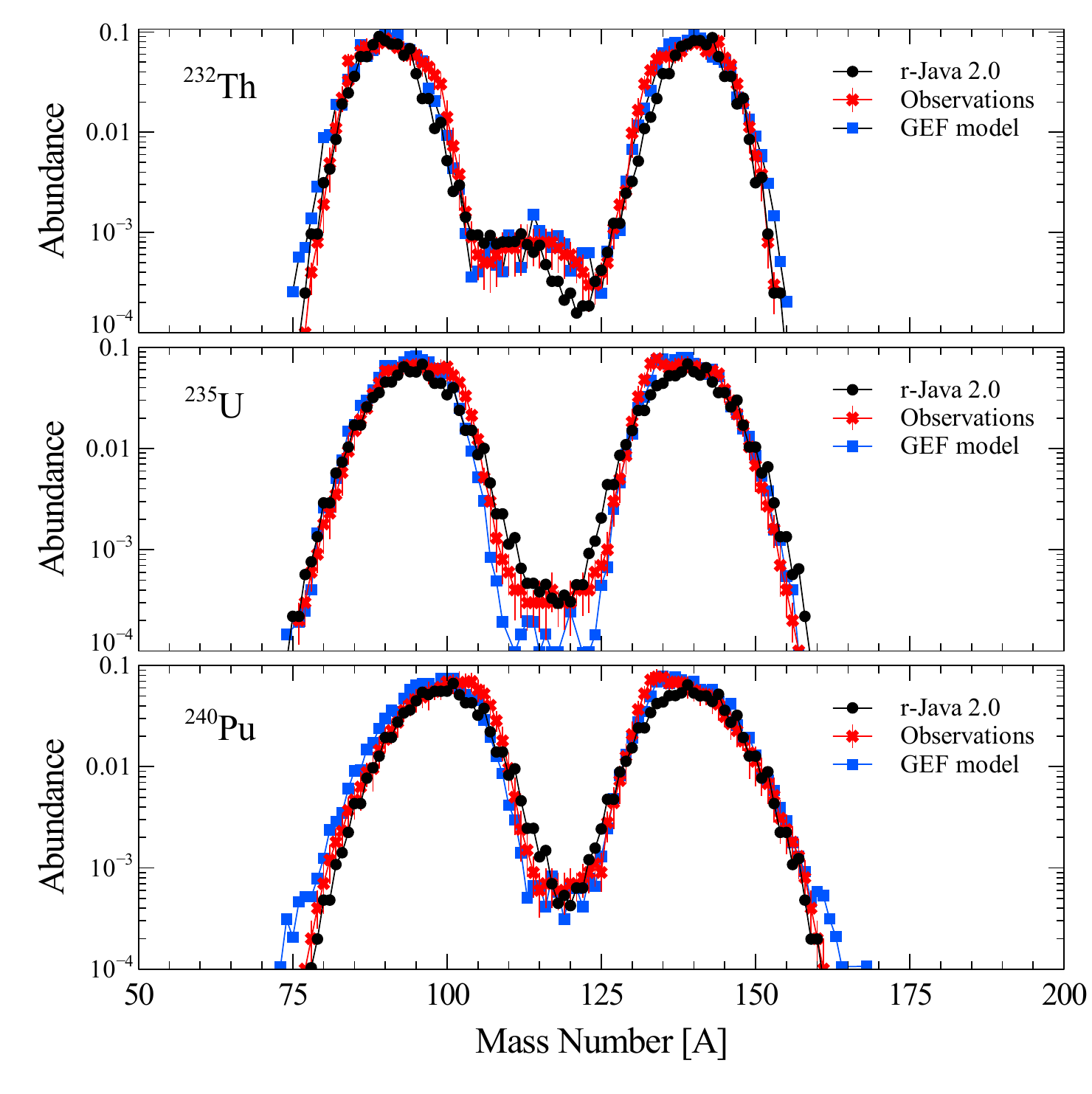}
\caption{\textit{Top:} The fission fragment mass distribution resulting for neutron-induced fission of $^{232}$Th by 1.5 MeV neutrons.\textit{Middle:} The fission fragment mass distribution resulting for neutron-induced fission of $^{235}$U by 1.5 MeV neutrons.  \textit{Bottom:} The fission fragment mass distribution resulting for neutron-induced fission of $^{240}$Pu by 1.5 MeV neutrons.  }
\label{fissfrag}
\end{figure*}

In r-Java 2.0 for the full fission treatment three mass fragmentation channels are considered and neutron evaporation is explicitly handled for each fission event.  The probability that the fission will follow a symmetric scission or one of the two standard channels is determined by integrals over the level density up to the available energy at the saddle point \citep{ben98,schmidt10}.  The first standard channel (SI) results in the heavier fission fragment containing 82 neutrons and for the second standard channel (SII) the heavier fission fragment contains approximately 88 neutrons.  The likelihood of the fission event following a particular standard channel is parameterized by the relative strength ($CI$ and $CII$) and depth ($\delta VI$ and $\delta VII$) of the corresponding valleys in the potential energy landscape at scission.  For r-Java 2.0 the strength and depth of the standard channel parameters are found through fitting observed fission fragmentation distributions for a range of nuclei between $^{232}$Th and $^{248}$Cm \citep{chadwickFiss}.  The remaining fissile and fertile nuclei use the standard channel parameter values of $^{235}$U as the default values, however these parameters can be adjusted using the fission module of r-Java 2.0.  The mass fragmentation distributions for $^{232}$Th, $^{235}$U and $^{240}$Pu are displayed in Fig. \ref{fissfrag}.  For reference in Fig. \ref{fissfrag} the results calculated using the fission module in r-Java 2.0 are compared to the results of the GEF model \citep{schmidt10} as well as observations \citep{chadwickFiss}. 

The result of the mass fragmentation calculation for each fissionable parent is that a probability distribution of potential daughter pairs is found.  In r-Java 2.0 the probability for each daughter species is multiplied by the parent fission rate and is incorporated as the daughter production rate in the network calculation. 

\begin{figure*}
\includegraphics[width = \linewidth]{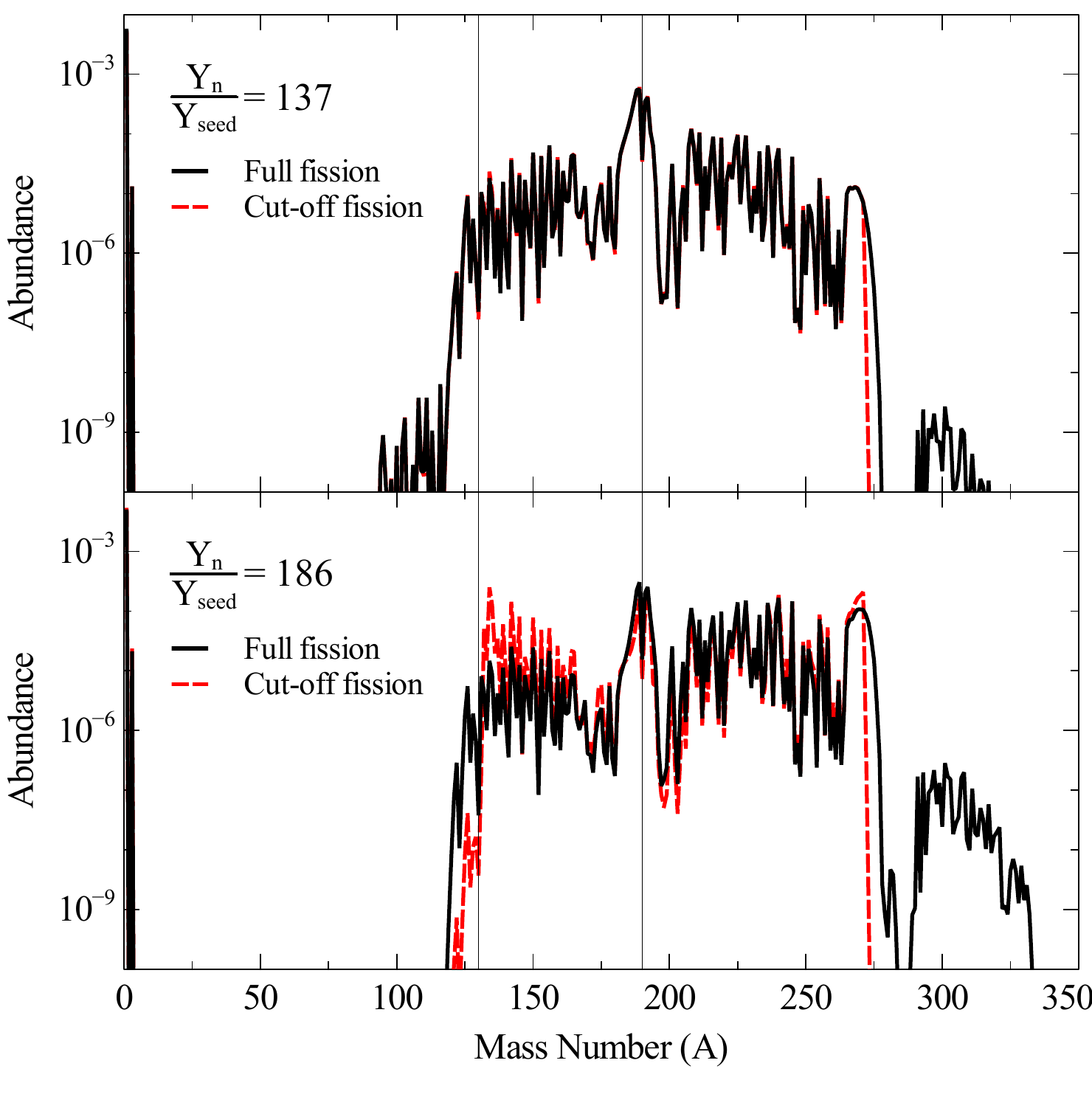}
\caption{Displayed is a comparison of the full fission methodology to the mass cut-off approach.  Two different initial neutron-to-seed ratios (\textit{top:} $Y_{\rm n}/Y_{\rm seed}\sim 137$,  \textit{bottom:} $Y_{\rm n}/Y_{\rm seed}\sim 186$) are considered while all other parameters remained the same (see section \ref{fiss} of text for details).  In both panels the red dashed line denotes the final abundance of a simulation that used the mass cut-off approach while the black solid line represents the full fission treatment.  The relevant magic numbers are highlighted with a fine vertical black line.  }
\label{fissDist}
\end{figure*}

A comparison of r-process final abundance distributions for both the mass cut-off and full fission treatment can be seen in Fig. \ref{fissDist}.   For the cut-off fission methodology a maximum mass of A$\,= \,$272 was used and each fissioning nuclei splits into two daughter species.  The full fission treatment uses the three fission processes discussed above as well as the fission fragmentation calculation.  For each simulation displayed in Fig. \ref{fissDist} the same initial abundance of iron-group nuclei were used, starting from the same initial temperature of $1.0 \times 10^9$K.  The initial mass density was $10^{11}$g cm$^{-3}$ for each simulation run which followed the same density profile, $\rho(t) = \rho_0 / \left(1+1.5\,t/\tau\right)^2$ with an expansion timescale ($\tau$) of 0.003 s.  The only variation between simulation runs shown in the top and bottom panels of Fig. \ref{fissDist} is the neutron-to-seed ratio ($Y_{\rm n}/Y_{\rm seed}$).  For the top panel $Y_{\rm n}/Y_{\rm seed}$= 137 was used and the bottom panel displays the r-process yield of a more neutron-rich simulation run which began with $Y_{\rm n}/Y_{\rm seed}$ = 186.  These two parameter sets were chosen to highlight the differences between the two fission methodologies. 

In the smaller $Y_{\rm n}/Y_{\rm seed}$ scenario, shown in the top panel of Fig. \ref{fissDist}, the r-process is just capable of breaking through the N = 184 magic number.  In this case the full fission run has been able to crossover a region of instability at about A$\,\sim\,$280 and has produced a small peak of super-heavies at about A$\,\sim\,$290.  Aside from the small super-heavy peak the results of both the full fission and cut-off methodology are largely the same.

For the larger $Y_{\rm n}/Y_{\rm seed}$ scenario seen in the bottom panel of Fig. \ref{fissDist}, the fission cut-off approach over-produces nuclei at A$\,\sim\,$130 by over ten times compared to the full fission treatment. This overproduction is due to the increased fission recycling caused by forcing all nuclei heavier than A$\,\sim\,$272 to undergo fission.  For this $Y_{\rm n}/Y_{\rm seed}$ the full fission simulation run produces a super-heavy peak on the order of $10^{-7}$.  The results of these simulation runs do not speak to the long-term stability of the super-heavy nuclei produced but rather shows the large variation between the two fission methodologies at the point of neutron freeze-out, which is to say when the neutron to r-process product ratio drops below one ($Y_{\rm n}/Y_{\rm r} < 1$).  

\begin{figure*}
\includegraphics[width = \linewidth]{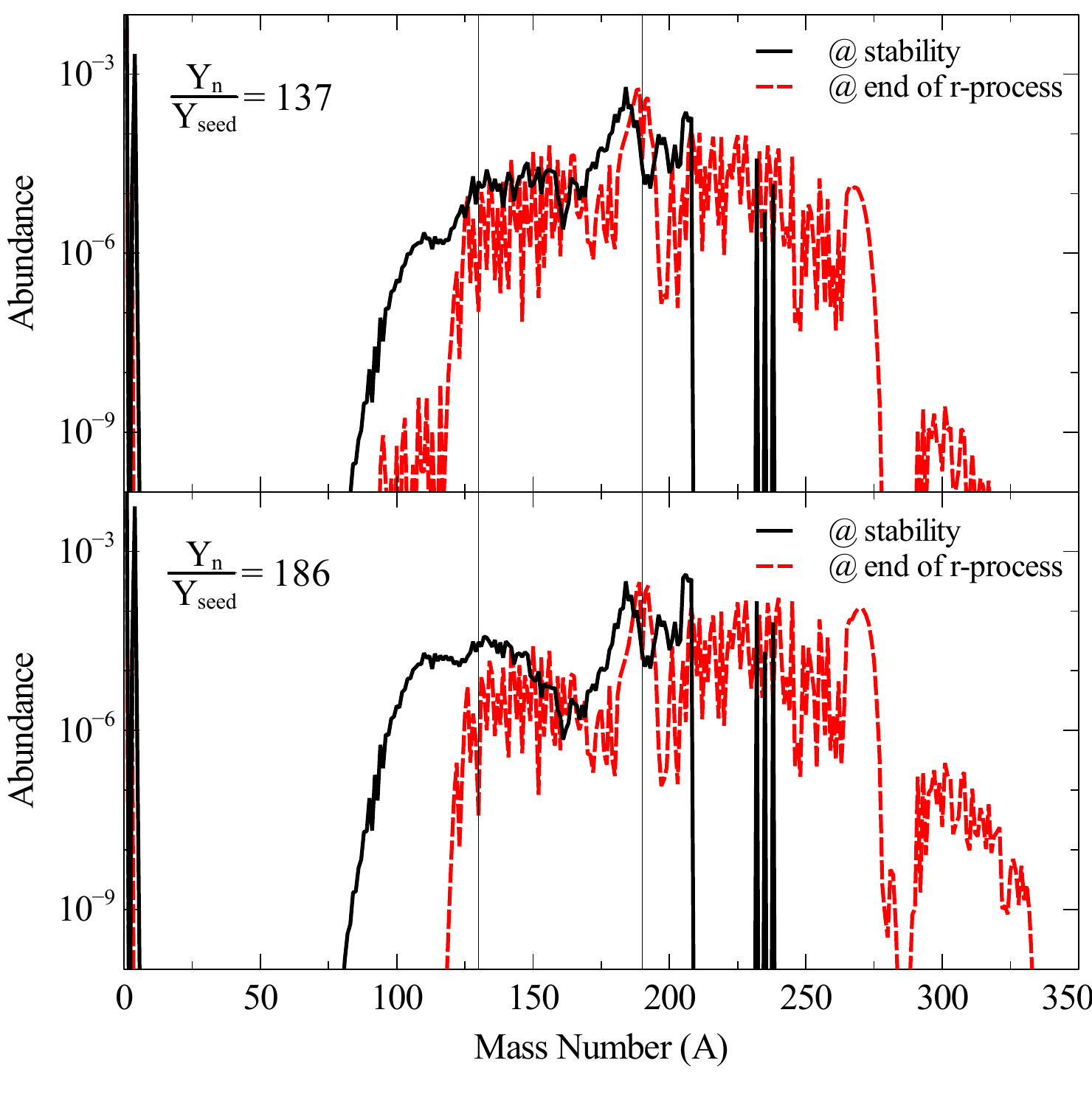}
\caption{Displayed is an overlay of the abundances of the same two initial neutron-to-seed ratios simulations shown in Fig. \ref{fissDist} ($Y_{\rm n}/Y_{\rm seed} = 186$ denoted by the black solid line and $Y_{\rm n}/Y_{\rm seed} = 137$ denoted by the red solid line) after having allowed the system to decay back to stability.  
For reference the abundances at the end of the r-process are included ($Y_{\rm n}/Y_{\rm seed} = 186$ denoted by the grey dotted 
line and $Y_{\rm n}/Y_{\rm seed} = 137$ denoted by the light red dashed line.)  The relevant magic numbers are highlighted with a fine vertical black line.}
\label{runOff}
\end{figure*}

The final abundances once the systems are allowed to decay to stability can be seen in Fig. \ref{runOff}.  Fission recycling gives rise to nearly all the nuclei abundances below A$\sim 150$ seen in both cases.  The distribution of fission recycled nuclei is similar in both cases because fission is occurring from the same region.  The shape of the fission contribution found using r-Java 2.0 coincides with the findings of \cite{peter08} who used the statistical code ALBA to calculated the fission yield for each fission event.  For comparison, the abundances at the moment r-process stops is included in Fig. \ref{runOff} for both neutron-to-seed simulation runs.  
The robust fission calculations included in r-Java 2.0 provides an accurate assessment of the role of fission recycling in the r-process.

\section{Beta-delayed Neutron Emission}
\label{bdn}
\begin{figure*}
\includegraphics[width = \linewidth]{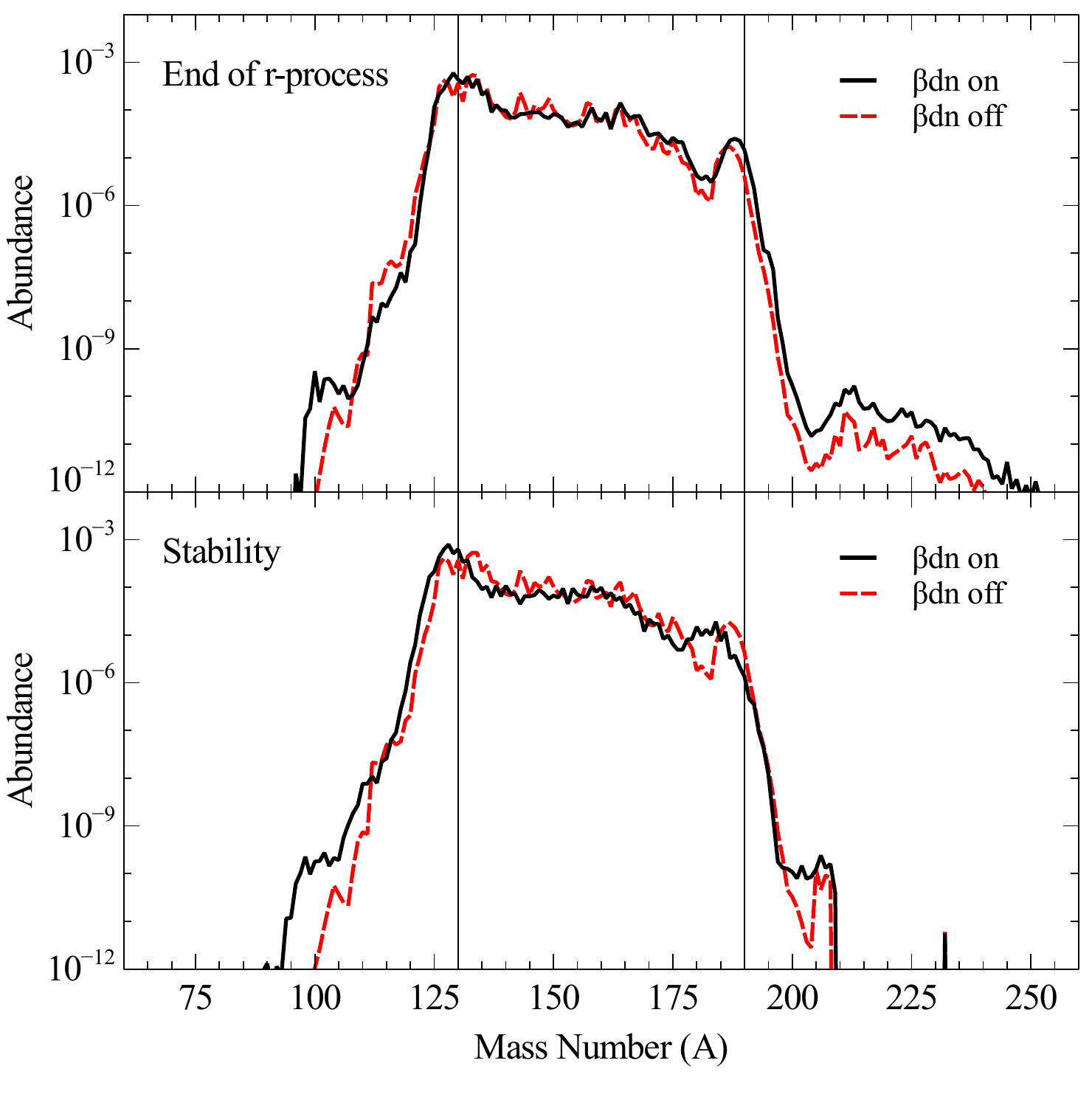}
\caption{The effect of $\beta$-delayed neutron emission on nuclei  abundance is compared, the black line denoting an r-process simulation that included $\beta$-delayed neutron emission and for the red dashed line that process was omitted.  The results plotted in this figure as well as in Figs.\ref{bdnNNevo} and \ref{bdnZvsN} are from simulation runs that were identical with the exception of whether or not $\beta$-delayed neutron emission was included. \textit{Top:} The nuclei abundances at the moment the neutron-to-seed ratio drops below one.  \textit{Bottom:} The nuclei abundances after decay to stability.  The relevant magic numbers are highlighted with a fine vertical black line.}
\label{bdnYvsA}
\end{figure*}

In order to study the effects of $\beta$-delayed neutrons on the r-process we compare two simulation runs that are identical except whether or not $\beta$-delayed neutron emission is included.  The top panel of Fig. \ref{bdnYvsA} displays a comparison of the abundance distributions at the end of the r-process, which for this study was defined to be once the neutron-to-r-process products ratio ($Y_{\rm n}/Y_{\rm r}$) drops below one.  The emission of $\beta$-delayed neutrons acts to smooth out the variability in nuclei distribution compared to that of the case without $\beta$-delayed neutrons.  The peak at A$\sim$188 is shifted slightly heavier with the inclusion of $\beta$-delayed neutrons and as well the abundance of nuclei with mass greater than A = 200 is increased.  The lower panel of Fig. \ref{bdnYvsA} shows the final nuclei abundance distribution once the systems are allowed to decay to stability.  For the simulation that did not include $\beta$-delayed neutron emission the nuclei abundance distribution below A $\simeq$209 remains virtually unchanged from the time r-process stops to that of stability.  However when $\beta$-delayed neutrons are included the decay to stability causes further reduction in the variability of the abundance distribution and a shifting of the peaks towards lower mass.

\begin{figure*}
\includegraphics[width = \linewidth]{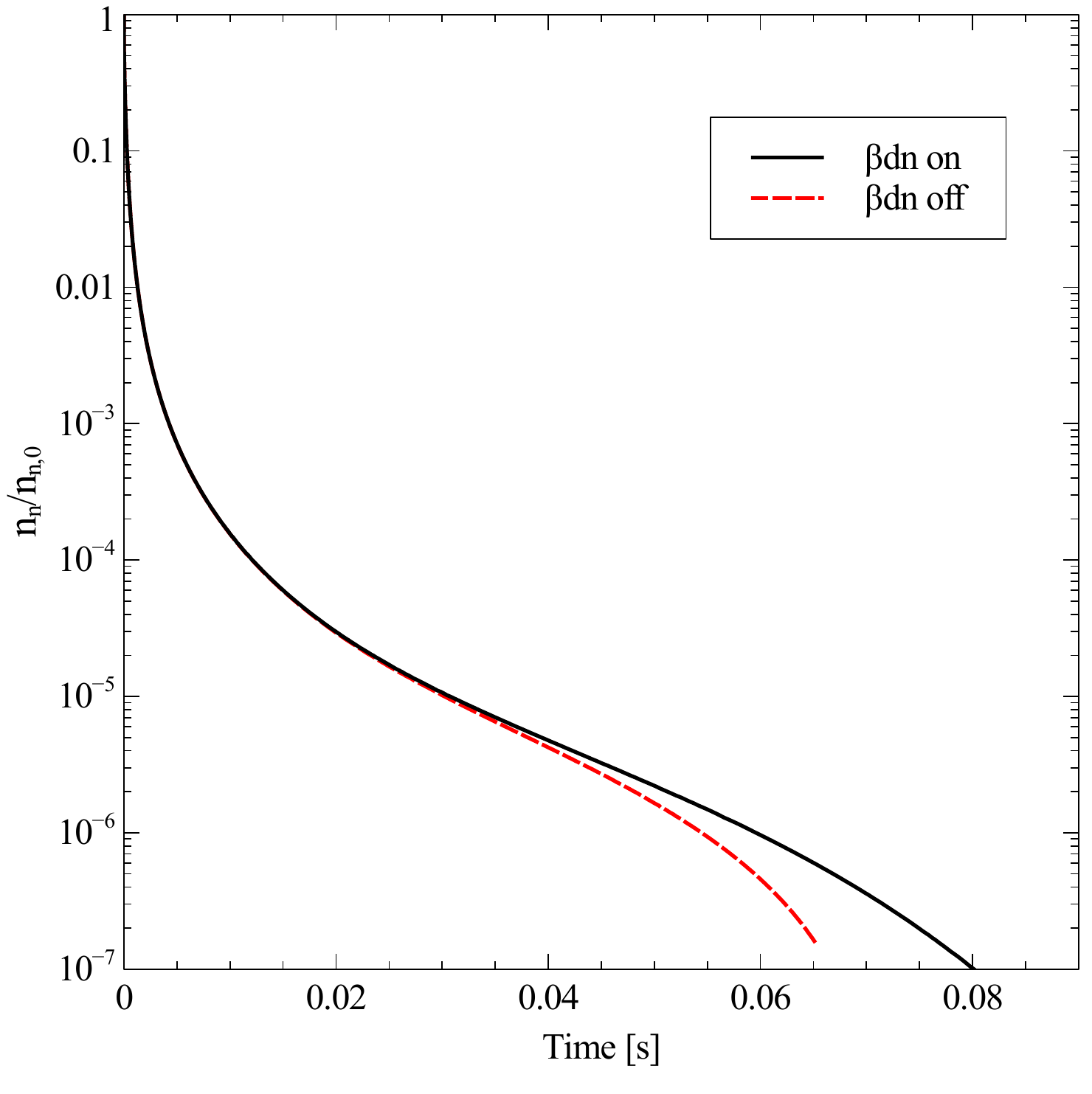}
\caption{The evolution of neutron density until the r-process is terminated is plotted.  The black line denotes an r-process simulation that included $\beta$-delayed neutron emission and for the red dashed line that process was omitted.  }
\label{bdnNNevo}
\end{figure*}

In Fig. \ref{bdnNNevo} the evolution of neutron density during the r-process is compared between the simulations with and without $\beta$-delayed neutron emission.  The $\beta$-delayed neutrons act to keep the neutron density higher for longer when compared to the case in which $\beta$-delayed neutron emission was ignored.  By bolstering the neutron density, $\beta$-delayed neutron emission allows the r-process to proceed more readily to heavier elements, an effect that can can be seen in the top panel of Fig. \ref{bdnYvsA}.

\begin{figure*}
\includegraphics{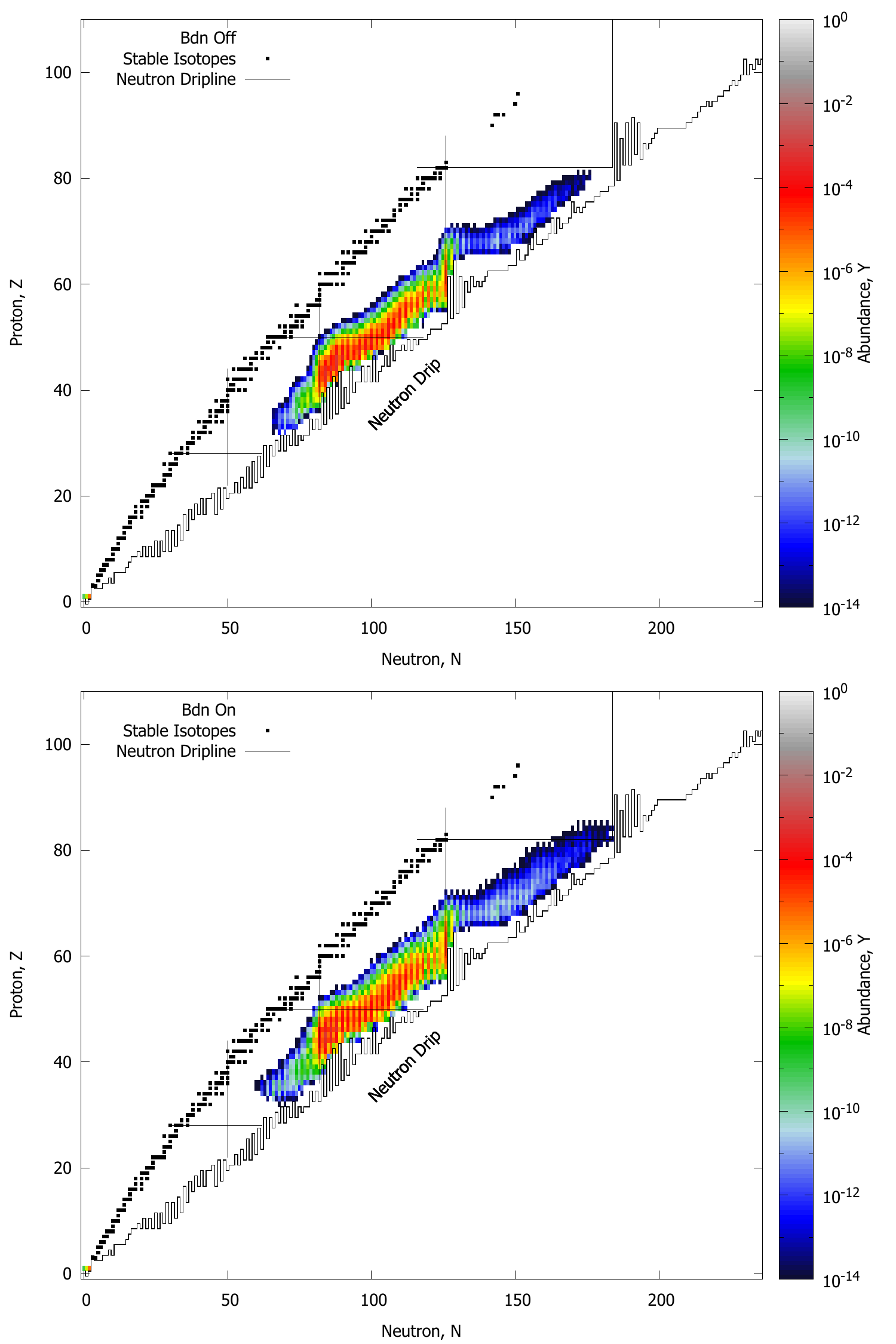}
\caption{ The nuclei abundances at the moment the neutron-to-seed ratio drops below one is plotted on the (N,Z) plane.  Stable nuclei, the location of the proton and neutron closed shells and the neutron drip line are included for reference.  \textit{Top:} Simulation that did not include $\beta$-delayed neutron emission.  \textit{Bottom:} Simulation including $\beta$-delayed neutron emission.}
\label{bdnZvsN}
\end{figure*}

The abundances of nuclei at the end of the r-process plotted on the (N,Z) plane can be seen in Fig. \ref{bdnZvsN} (top panel displays the case where $\beta$-delayed neutron emission was ignored and the bottom panel the case with its inclusion).   For the simulation run that included $\beta$-delayed neutrons the r-process accesses a broader  range (along lines of constant Z) of nuclei, reaching closer to the valley of stability.  This broadening effect caused by $\beta$-delayed neutrons is most noticeable around the N = 82 and 126 closed shells.  The ability of $\beta$-delayed neutron emission to allow for matter-flow past the N = 126 closed shell can be seen in Fig. \ref{bdnZvsN} as the breadth of populated nuclei and abundance in the region past N =126 is increased in the case in which $\beta$-delayed neutron emission is included.

\section{Full Reaction Network}
\label{full}
In order to expand beyond the WPA, reactions that stay within an isotopic chain, namely neutron-capture and photo-dissociation, must be included in the network calculation.  This means that rather than solving a system of equations the size of which is determined by the number of isotopic chains ($110$) as in the WPA case, for the full network case an equation for every nuclei must be included (a total of $8055$).  The computational cost of this addition is significant since finding a solution to a reaction network scales as $N^3$ where $N$ is the number of coupled differential equations.  However there are methods that can be invoked to mitigate this cost; we take advantage of the fact that each nuclei in the network is only coupled to another nuclei if there is an adjoining reaction (i.e. nuclei (Z,A) is coupled to both (Z+2,A+4) and (Z-2,A-4) via $\alpha$ decay).  This is effectively utilizing the sparseness of the reaction rate matrix which alleviates memory load issues and speeds up runtime.  We solve the fully implicit network using the Crank-Nicholson method.  The rate of thermonuclear energy released (or absorbed) is calculated using the methodology laid out by \cite{hix06}.

\begin{figure*}
\includegraphics[width = \linewidth]{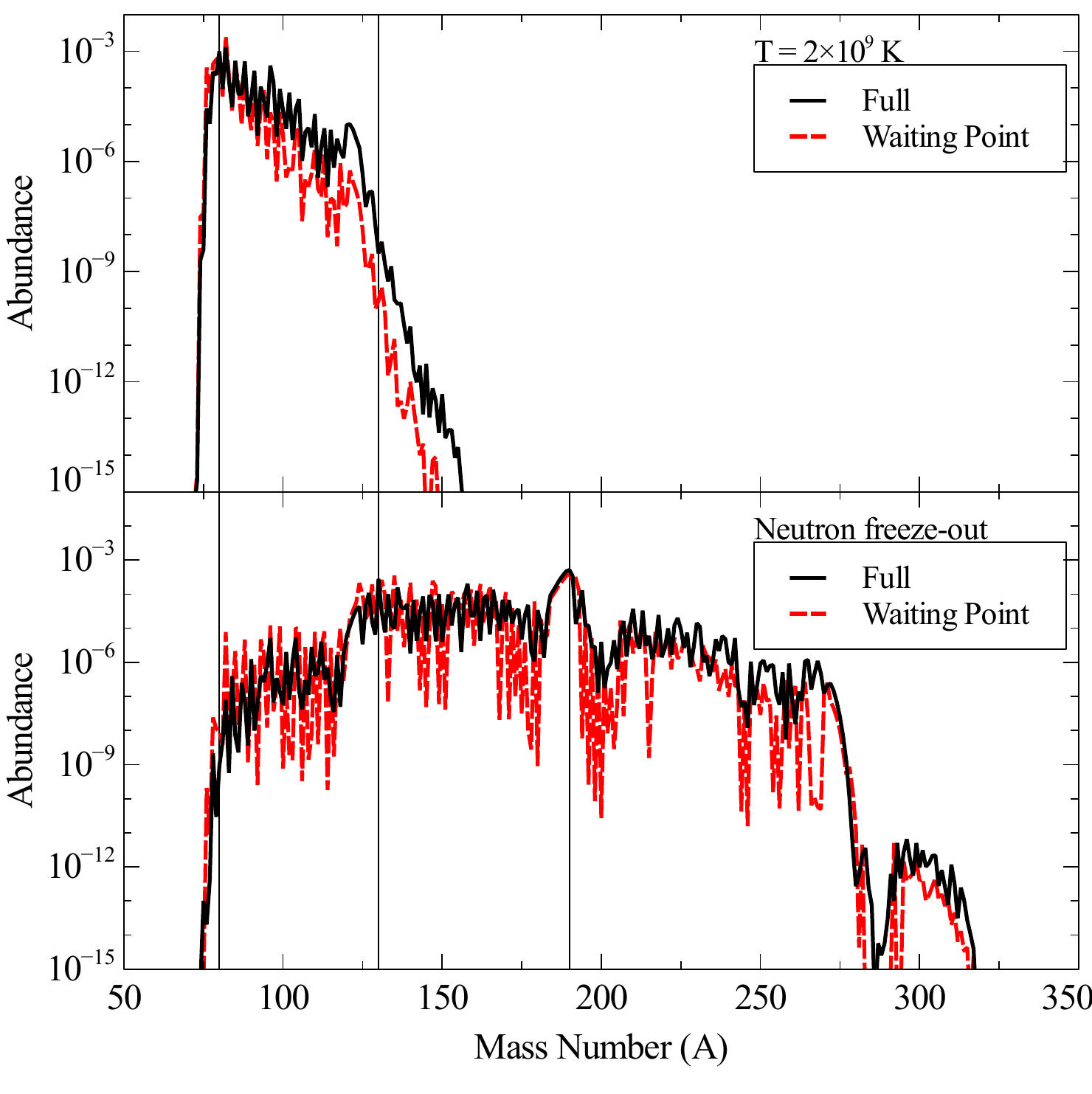}
\caption{Comparison of the simulation results from the WPA (red dashed line) with that of the full network (black solid line). \textit{Top:} Nuclei abundances when the temperature drops to $2\times 10^9$K. \textit{Bottom:} The nuclei abundances at neutron freeze-out, see text for details of stopping criteria.  The relevant magic numbers are highlighted with a fine vertical black line.}
\label{wpVSfull}
\end{figure*}

Fig. \ref{wpVSfull} highlights the importance of the imposed stopping criteria on network calculations through a comparison of the results from the WPA to that of the full network.  For the results plotted in both panels of Fig. \ref{wpVSfull} the same initial conditions and expansion profiles\footnote{$\rho(t) = \rho_0/ \left(1 + t/0.001\right)^2$} were chosen.  The simulations considered begin from an iron seed with; Y$_{\rm e,0}= 0.16$, T$_{0}= 4 \times10^9$ K and $\rho_0=10^{10}$g cm$^{-3}$).  In the top panel of Fig. \ref{wpVSfull} the calculations are stopped when the temperature falls below $2\times 10^9$ K, an imposed cut-off based on the work of \cite{cowan83}.  The nuclei distribution in the WPA simulation is peaked at A = 80 with lower abundances of nuclei up to A $\sim$ 120 and then a precipitous drop in abundance for heavier nuclei.  In the case where the full reaction network calculation was stopped once the temperature fell to $2\times10^9$ K (displayed in the top panel of Fig. \ref{wpVSfull}) there is good agreement to the WPA calculation.  The shape of the nuclei abundance distribution is the same for both network calculations, with the full reaction network producing a slightly greater abundance of heavy nuclei.  However the results displayed in the top panel of Fig. \ref{wpVSfull} are not indicative of the \textit{full potential} of the r-process for this chosen environment, since as the temperature drops below the imposed minimum cut-off the neutron density still remains high ($n_{\rm n} \sim 10^{30}$ cm$^{-3}$).  For the bottom panel of Fig. \ref{wpVSfull} the minimum temperature stopping criterion for the WPA was lowered to $10^9$ K and for this case both the WPA and full reaction network calculations halt at neutron freeze-out ($Y_{\rm n}/Y_{\rm r}$ = 1).  Once again both network calculations display similar nuclei abundance distributions.  The results of the WPA reflect the (n, $\gamma$)$\leftrightharpoons$($\gamma$, n) equilibrium which is not as accurate as the full treatment.  This is manifested as deeper troughs in nuclei abundance, especially around the A = 190 peak, and greater variability for the lower mass nuclei.  The smoother distribution in the full reaction network results is also due to the inclusion of $\beta$-delayed neutron emission.   

The users of r-Java 2.0 are afforded the option to choose the stopping criteria for r-process calculations; minimum temperature and neutron density for the WPA network and $Y_{\rm n}/Y_{\rm r}$ for both networks.

\section{Test Cases}
\label{bench}

As part of the testing phase of the development of r-Java 2.0 we  attempted to reproduce the results from three other full network r-process codes; the Clemson University nucleosynthesis code \citep{clemCode} which will furthermore be referred to as the Clemson code, the Universitat Basel nucleosynthesis code \citep{fried99}, to be referred to as the Basel code and the nucleosynthesis code developed at Universit\'e Libre de Bruxelles \citep{gori11} which will be called the Bruxelles code for the remainder of this article.  While a complete \textit{apples to apples} comparison was not tenable, the results of our tests showed good agreement with each of the three codes studied.  

\subsection{Clemson Nucleosynthesis Code}
\label{clem}
\begin{figure*}
\includegraphics[width = \linewidth]{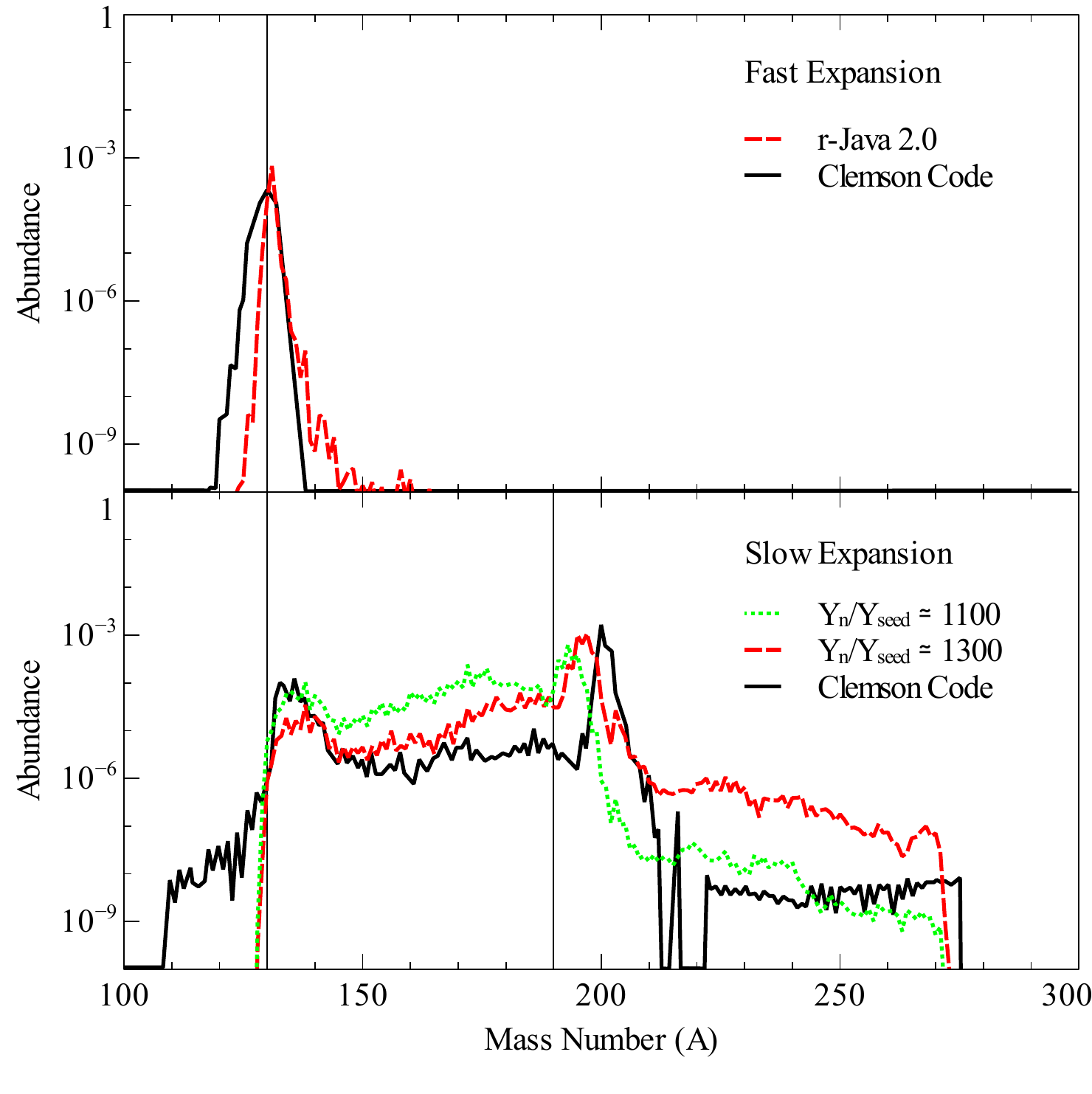}
\caption{\textit{Top:} A comparison of the final abundances from r-Java 2.0 (red dashed line) and the Clemson nucleosynthesis code (black solid line) for a fast expansion r-process site.  \textit{Bottom:} A comparison of the final abundances from r-Java 2.0 with two different initial neutron-to-seed ratios ($Y_{\rm n}/Y_{\rm seed} \sim 1100$ denoted by the green dotted line, $Y_{\rm n}/Y_{\rm seed} \sim 1300$ by the red dashed line) and the Clemson nucleosynthesis code (black solid line) for a slow expansion r-process site.  The relevant magic numbers are highlighted with a fine vertical black line.}
\label{clemComp}
\end{figure*}

For the Clemson code comparison, seen in Fig. \ref{clemComp}, we endeavoured to reproduce the results shown in Fig. 7 and 8 of \cite{jaik07}.  We found that the initial abundance was not of much importance for either case because the neutron-to-seed ratio was high enough such that any influence from the initial abundance was washed away by the r-process.  The top panel of Fig. \ref{clemComp} shows the results of a fast expansion r-process site and the bottom panel a slow expansion (corresponding to Fig.7 and 8 from \cite{jaik07} respectively).  In the fast expansion case both r-Java 2.0 and the Clemson code show that the r-process is not capable of proceeding past the A = 130 magic number.  In the slow expansion case the environment remains favourable for r-process much longer and the final abundance for both r-Java 2.0 and the Clemson code contains peaks shifted to the heavy-side of the A=130 and A=190 observed solar peaks.  The differences between the final abundances from r-Java 2.0 and the Clemson code seen in both cases can be credited to the fact that the two codes use different mass models (the Clemson code used the finite range droplet model and r-Java 2.0 HFB21) which has been shown to affect the r-process abundance yield \citep[e.g.][]{farouqi10}.

\subsection{Basel Nucleosynthesis Code}
\label{basel}
In order to compare to an updated version of the Basel code we pushed to reproduce the abundances shown in Fig. 10 of \cite{farouqi10} which considers the HFB17 mass model.  As described in \cite{farouqi10} the r-process network begins at the termination of the charged particle network, thus we used the abundance per mass number at the end of the charged-particle network displayed in Fig. 5 of \cite{farouqi10} to determine our initial seed nuclei distribution for comparison.  Having only the abundance per mass number information we had to choose which nuclei to set each abundance to in order to build our initial seed nuclei.  We made the assumption that the system is in (n,$\gamma$)$\leftrightharpoons$($\gamma$,n) equilibrium at the beginning of the r-process based on the initial conditions used in \cite{farouqi10} of $T = 3 \times 10^9$ K and $n_n = 10^{27}$ cm$^{-3}$.  Then for abundance at each mass number plotted in fig. 5 of \cite{farouqi10} we set it to the isotope which most closely matched the predictions of the nuclear Saha equation.  For each different entropy simulation that we ran these abundances were uniformly scaled such that they produced the correct seed abundance as shown in Fig.3 of \cite{farouqi10}.  The initial neutron abundance was then determined from the neutron-to-seed ratio stated in Table 5 of \cite{farouqi10}.  The use of the same initial abundance distribution for each simulation run by r-Java 2.0 which may not have been the case for the simulations done by \cite{farouqi10}, is the largest potential source of discrepancy in this comparative analysis.

Consistent with \cite{farouqi10} we used Y$_{\rm e}$ = 0.45 and started our simulations with an initial temperature of $3\times10^9$K.   We followed the same constant entropy methodology described in \cite{farouqi10} to evolve temperature and density.  In this scenario the temperature evolves adiabatically and the entropy is assumed to be radiation dominated which allows for the inference of the evolution of matter density.  The time-dependence of the temperature and matter density ($\rho_5$ is in units of $10^5$ g cm$^{-3}$) are thus governed by the following equations,

\begin{equation}
T_9 \left( t \right) = T_9 \left( t=0 \right)\frac{R_0}{R_0 + v_{\rm exp}\,t} \ ,
\end{equation}
\begin{equation}
\rho_5 \left( t \right) = 1.21 \frac{T_9^3}{S}\left(1+\frac{7}{4}\frac{T_9^2}{\left(T_9^2 + 5.3\right)}\right) \ ,
\end{equation}

\noindent where $R_0 = 130$ km and $v_{\rm exp} = 7500$ km s$^{-1}$.

\begin{figure*}
\includegraphics[width = \linewidth]{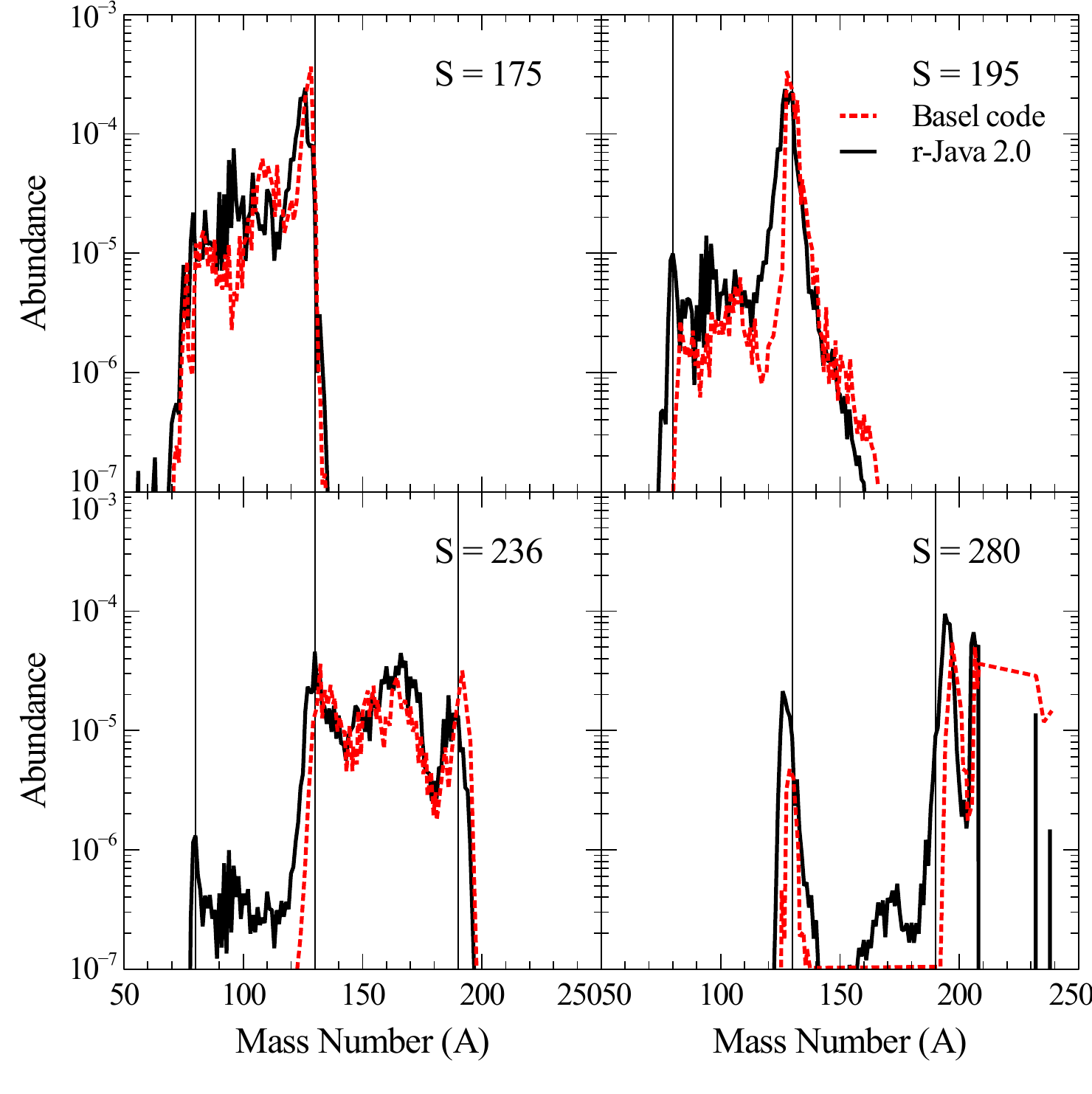}
\caption{A comparison of r-process abundance yields as calculated by r-Java 2.0 (black solid line) and the Basel nucleosynthesis code (red dashed line).  The relevant magic numbers are highlighted with a fine vertical black line.  For each panel a different entropy was assumed, which changes the initial neutron-to-seed ratio as well as the evolution of the density, see text for details.   \textit{Top-Left:} Simulation run assuming the entropy of the wind is S = 175.   \textit{Top-Right:} Simulation run assuming the entropy of the wind is S = 195.   \textit{Bottom-Left:} Simulation run assuming the entropy of the wind is S = 236.  \textit{Bottom-Right:} Simulation run assuming the entropy of the wind is S = 280.  See text for details of initial conditions.  }
\label{baselComp}
\end{figure*}

In order to maintain consistency with the Basel code we terminated the r-process once the neutron-to-seed ratio dropped below one and the abundances shown in Fig. \ref{baselComp} are after decay back to stability. 

The top-left panel of Fig. \ref{baselComp} which displays the results of the S = 175 simulation runs, shows the best agreement between r-Java 2.0 and the Basel code of all the cases tested.  Both r-Java 2.0 and the Basel code show a final nuclei abundance that predominantly ranges from $70<$A$<$135.  The results from each code displays a peak below the A = 130 magic number, however the Basel code peak is shifted heavier with respect to that of r-Java 2.0. 

 The S = 195 simulation results (displayed in the top-right panel of Fig. \ref{baselComp}) from the Basel code and r-Java 2.0 are both dominated by a peak at the A =130 magic number.  The differences between the final abundances from the two codes for this entropy are consistent with differing initial abundances.  The fact that the results from r-Java 2.0 display a more distinct peak at the A = 80 magic number is consistent with the simulation run of r-Java 2.0 starting with more nuclei below the A = 80 magic number.  This would lead to nuclei piling up at A = 80 for r-Java 2.0 which would not be the case for the Basel code simulation run.  The difference in initial abundance also has an effect for the heavy side of the final abundance distribution.  With more nuclei initially between the A = 80 and A = 130 observed solar peaks the r-process simulation run of the Basel code is more capable of pushing through the A = 130 magic number to higher masses.  As for the r-Java 2.0 simulation once the r-process pushes through the A = 80 magic number nuclei will pile up on the light-side of the A = 130 magic number.  By the time the r-process reaches the A = 130 peak in the r-Java 2.0 run the neutron density will have dropped too low to significantly push past the A = 130 magic number.  The result of this is the increase production of nuclei on the lower mass side of the A =130 peak for the r-Java 2.0 simulation run with respect to that of the Basel code and a larger high mass tail in the Basel code simulation. 
 
Similar to the S = 195 case, the presence of nuclei below the A = 80 magic in the S =236 r-Java 2.0 simulation (seen in the lower-left panel of Fig. \ref{baselComp}) leads to the final abundance containing nuclei around A = 80 which is not the case for the Basel code results.  Once again this difference is consistent with different initial abundances for the two runs.  Neglecting the relatively small abundance for  $80\lesssim$A$\lesssim$125 in the r-Java 2.0 results, the final distributions of the S = 236 simulation runs for both codes are consistent with peaks around A =80, 165 and 190.  The A = 190 peak in the Basel code simulation is stronger and shifted towards heavier masses with respect to that of r-Java 2.0 which can be attributed to fact that the r-Java 2.0 simulation had more nuclei \textit{stuck} below the A = 80 magic number.  

The S = 280 simulation runs seen in the lower-right panel of Fig. \ref{baselComp} shows the same basic features for both codes.  The final nuclei abundance for both codes contains strong peaks at A =130 and A = 195, with the r-Java 2.0 results displaying stronger peaks.  The increased abundance of Th and U at stability in the Basel code simulation run could be due to the initial abundance differences discussed for the S = 195 and S = 236 cases or due to different definitions of stability.  For the r-Java 2.0 simulations the systems decayed for 13 Gyr or until the percent change in any nuclei abundance was less than $1 \times 10^{-15}$.

\subsection{Universit\'e Libre de Bruxelles Nucleosynthesis Code}
\label{brux}
\begin{figure*}
\includegraphics[width = \linewidth]{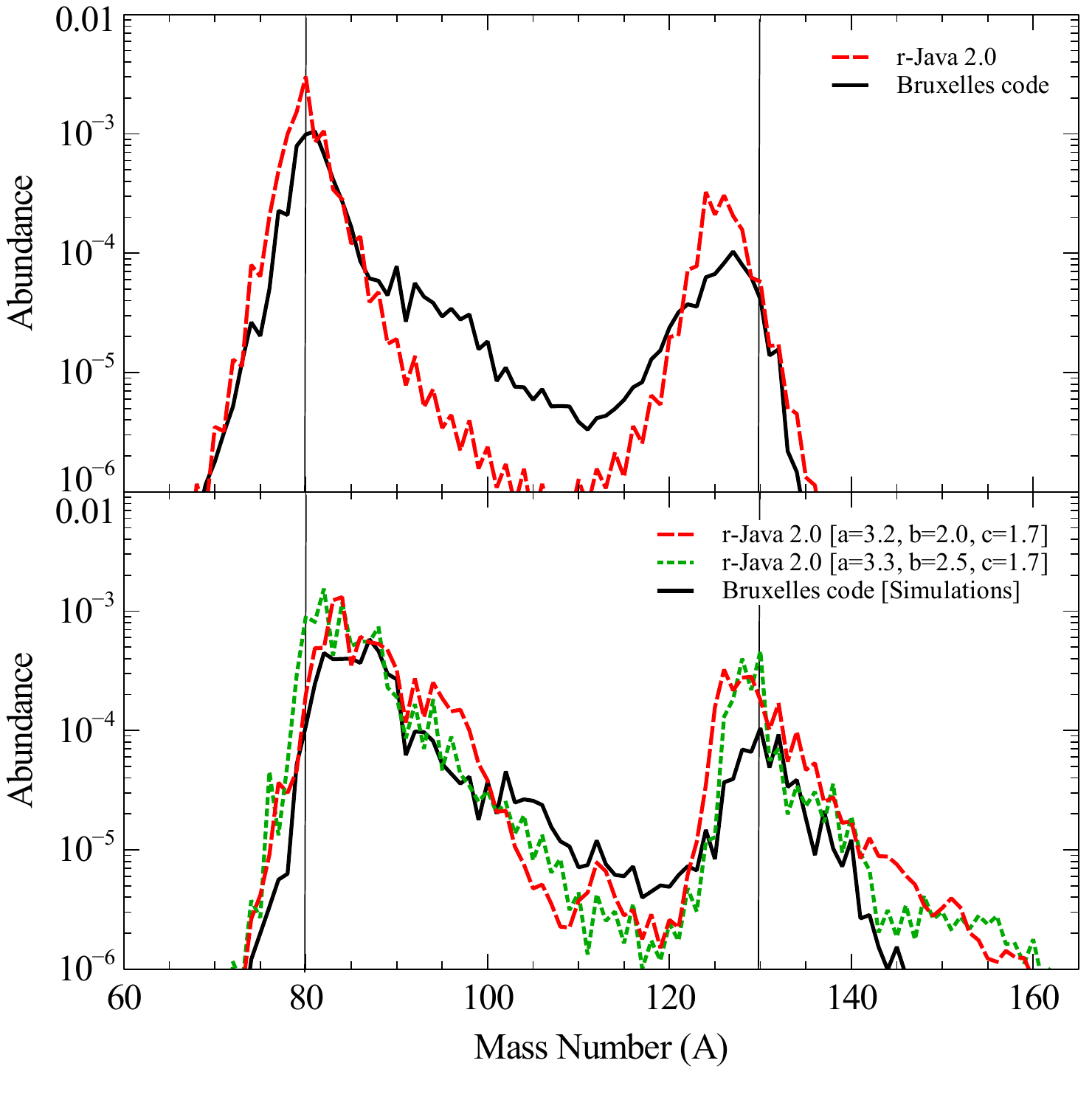}
\caption{\textit{Top:} The initial abundances used for the comparison of r-process simulations from r-Java 2.0 and the Bruxelles nucleosynthesis code.  The red dashed line denotes r-Java 2.0 and the black solid line the Bruxelles code.  \textit{Bottom:} The final abundances from r-Java 2.0 considering two different density evolution profiles (red dashed line and green dotted line) compared to that of the Bruxelles code (black solid line).  See text for details of simulations.  The relevant magic numbers are highlighted with a fine vertical black line. }
\label{bruxInit}
\end{figure*}

For our comparison to the Bruxelles code we attempted to reproduce the abundances after decompression displayed in Fig. 10 of \cite{gori11}.  As discussed in \cite{gori11} the initial abundances used for the r-process simulation are important because in this scenario the initial neutron-to-seed ratio is roughly 5 and the r-process is only capable of shifting the abundances toward heavier nuclei without dramatically altering the relative shape of the abundance distribution.  \cite{gori11} provides the initial abundances used for the r-process simulation which are calculated under NSE with Coulomb interactions included.  A comparison of the initial abundances of the Bruxelles code and r-Java 2.0 can be seen in the top panel of Fig. \ref{bruxInit}.  The peaks roughly centred at A = 80 and 125 as calculated by r-Java 2.0 are higher than that from the Bruxelles code, while the intermediate mass region is more abundant in the Bruxelles calculation.  The NSE calculation performed by r-Java 2.0 assumes Maxwell-Boltzmann statistics while the Bruxelles code used Fermi-Dirac which accounts for the differences in abundances.  While the nuclear physics used in the Bruxelles code is the most similar to r-Java 2.0 of all the codes studied, we had to implement an analytic approximation to the density evolution used by \cite{gori11}.  In order to compare to the Bruxelles code we chose the density profile shown in eqn. \ref{rho}
\begin{equation}
\label{rho}
\rho(t) = \rho_0 \left(\frac{1}{1+\left(a\,t/\tau\right)^b}\right)^c \ ,
\end{equation} 

\noindent where a, b and c are free parameters.  A value of $3\times10^{-4}$ seconds was used for the expansion timescale ($\tau$) which is consistent with that used by \cite{gori11}.

\begin{figure*}
\includegraphics[width = \linewidth]{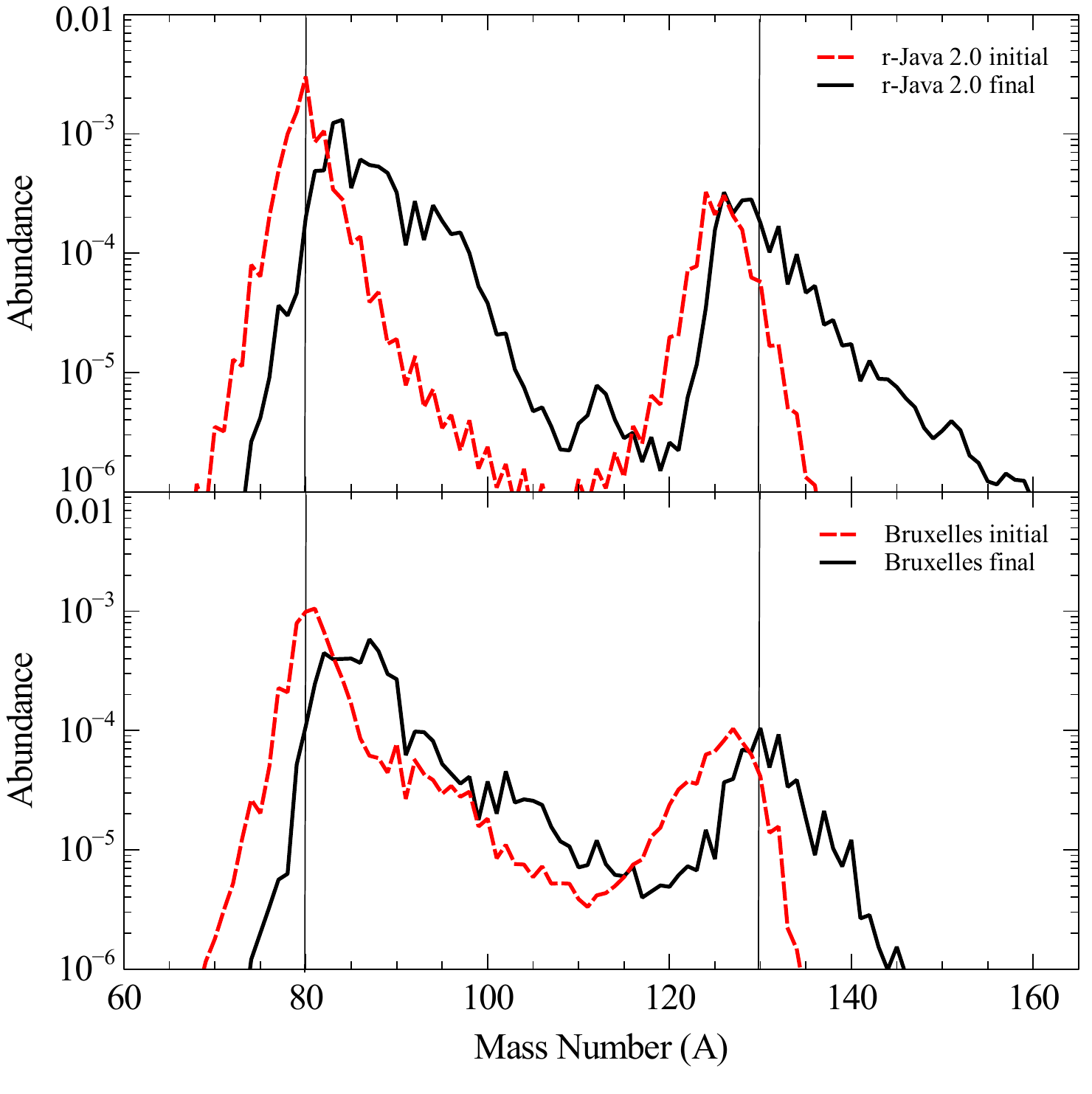}
\caption{ \textit{Top:} The final (black solid line) and initial (red dashed line) abundances as calculated by r-Java 2.0 for comparison to the Bruxelles code.  \textit{Bottom:} The final (black solid line) and initial (red dashed line) abundances as calculated by the Bruxelles code.  See text for details of simulations.  The relevant magic numbers are highlighted with a fine vertical black line.}
\label{bruxFinInit}
\end{figure*}

\noindent A comparison of two different sets of free parameters used in the density profile of r-Java 2.0 to the final abundances of the Bruxelles code can be seen in the bottom panel of Fig. \ref{bruxInit}.  As expected the differences in initial abundances are carried through to the final nuclei abundances with r-Java 2.0 displaying higher peaks at approximately A = 85 and 130 with the intermediate mass region more strongly produced in the Bruxelles code simulation.  To show that in both codes the r-process has the same effect on abundances in Fig. \ref{bruxFinInit} the final and initial abundances are overplotted for each code respectively.  For both r-Java 2.0 and the Bruxelles nucleosynthesis code r-process acts to shift the peaks towards heavier nuclei.

\section{Summary and Conclusions}
\label{summary}

This paper has discussed the nuclear physics incorporated in r-Java 2.0; providing cutting-edge fission calculations, $\beta$-delayed neutron emission of up to three neutrons and neutron capture and photo-dissociation rates from one of the most sophisticated mass models (HFB21).  Nevertheless it is the flexibility to change any parameter quickly and easily that makes r-Java 2.0 a powerful tool for the study of nuclear astrophysics.  As well r-Java 2.0 is capable of solving a full r-process reaction network containing over 8000 nuclei and can do so both accurately and efficiently, with a typical full reaction network simulation completed on the order of minutes.  The scientific aim of this release of r-Java is to study r-process nucleosynthesis in the expansion phase ($T \lesssim 3 \times 10^9$ K \cite[e.g.][]{howard93}) and NSE at high temperature ($T \gtrsim 4 \times 10^9$ K \cite[e.g.][]{tru66}).  We are currently developing a charged-particle reaction network module that will be incorporated in a future version of r-Java. 

With a more realistic treatment of fission we have added to r-Java 2.0 the ability to investigate the role of fission recycling in the r-process.  In the past by simply using the mass cut-off approach the mistake of going to too high of a neutron density was masked by the fact that fission recycling would not allow the r-process to proceed beyond the cut-off.  This manifests in the r-process abundance at neutron freeze-out in two ways; an under-production of super-heavy nuclei and the over-production of nuclei around the A=130 magic number.  With the fission methodology implemented here the super-heavy regime (A $>$ 270) can be studied using r-Java 2.0. The preliminary study undertaken here supports the findings of \cite{peter12}, where super-heavy nuclei (A$\sim 290$) can be formed by the r-process.  The super-heavies subsequently decay on the order of seconds.

The emission of $\beta$-delayed neutrons can act to maintain a sufficiently high neutron density to allow for the r-process to reach heavier elements.  The effect of $\beta$-delayed neutron emission is as well significant during the decay to stability once the r-process has stopped.  They act to smooth out the nuclei distribution on the path to stability and shifts the abundances to lower masses.  Their role may in some cases not be as direct as just stated.  The $\beta$-delayed neutrons can alter the r-process path, accessing nuclei which would more readily capture neutrons causing the neutron density to drop more rapidly than if they were ignored.  This must be studied in more detail and with r-Java 2.0 the user can quickly and easily investigate the effect of $\beta$-delayed neutrons on r-process abundances. 

By performing a comparative study between r-Java 2.0 and three other full network r-process codes we have found good agreement between the codes, however undertaking this analysis has highlighted the potential pitfalls of comparing the results from different codes.  Factors such as; choice of mass model, evolution methodology of physical parameters, code stopping criteria and precision can contribute to variations in r-process abundances that are artefacts of the nucleosynthesis code structure rather than physical scenarios being studied.  This comparative analysis highlights the universality of r-Java 2.0, which by allowing the user to customize both the nuclear and astrophysical parameters, is capable of reproducing the results of other nucleosynthesis codes.

The development of r-Java 2.0 was done in a way to maximize the flexibility of the software, allowing for the adjustment of any nuclear or physical property both quickly and easily.  The choice of Java as the programming language allowed for the inclusion of an easy to use GUI that is cross-platform compatible.  Beyond its applicability to scientific study, the goal of r-Java 2.0 was to make it accessible in a teaching capacity by ensuring it is easy to use and allowing for the investigation of individual processes.

In the follow-up paper to this work we will turn our attention to the astrophysical side of r-process which is well covered by r-Java 2.0.  Built into the interface of r-Java 2.0 is the option to define a custom density evolution or to select one of three proposed astrophysical r-process sites; high-entropy winds around proto-neutron stars \citep[example studies;][]{woos92,qianWoos96,thomp01,farouqi10}, ejecta from neutron star mergers \citep[][and others]{freid99b,gori11} or ejecta from quark novae \citep{jaik07}.  For each of the proposed astrophysical sites r-Java 2.0 consistently calculates the temperature and density evolution, the details of which will be discussed in the aforementioned upcoming paper.  By including the physics of different astrophysical sites in one piece of r-process software we have provided an even platform to compare the r-process abundances of different astrophysical sites.

\begin{acknowledgements}
   This work is supported by the Natural Sciences and Engineering Research Council of Canada.  NK acknowledges support from the Killam Trusts. 
\end{acknowledgements}

\bibliographystyle{aa}
\bibliography{paper2}

\clearpage
\begin{table*}
\caption{Description of symbols}              
\label{symbol}  
\centering                                   
\begin{tabular}{c l}        
\hline\hline                     
Symbol & Description \\    
\hline                                  
    Y$_0$(Z,A) & Initial abundance of isotope (Z,A)  \\   
    T$_0$ & Initial temperature \\
    $\rho_0$ & Initial mass density  \\
    $\rho$(t) & Density evolution profile  \\
    $\tau$ & Expansion timescale \\
    t$_{\rm sim}$ & Simulation duration \\
    Y$_{\rm e,0}$ & Initial electron fraction (WPA only) \\
    Z$_0$ & Initial element (WPA only) \\
\hline                                            
\end{tabular}
\end{table*}
\end{document}